\documentclass[10pt, journal]{IEEEtran}
\newcommand{\subparagraph}{}
\usepackage{cite}
\usepackage{amsmath,amssymb,amsfonts}
\usepackage{graphicx}
\usepackage{textcomp}
\usepackage[font=footnotesize]{caption}
\usepackage{comment}
\usepackage{subcaption}
\usepackage{physics,amsmath}
\usepackage[ruled,vlined]{algorithm2e}
\include{pythonlisting}
\usepackage{float}
\usepackage{graphics}
\usepackage{titlesec}
\usepackage{graphicx}
\usepackage{tabularx}
\usepackage{flushend}

\usepackage{xcolor}
\usepackage{titlesec}

\usepackage{graphicx}
\usepackage{mathtools}

\usepackage{array}

\usepackage[english]{babel}
\usepackage[utf8x]{inputenc}
\usepackage{xcolor}
\usepackage[colorinlistoftodos]{todonotes}

\usepackage{enumitem}

\usepackage{amsmath}
\usepackage{amssymb}
\usepackage{amsfonts}
\usepackage{amstext}
\usepackage{amsthm}
\usepackage{soul}

\usepackage{algpseudocode}
\usepackage{tikz}
\newcommand{\ballnumber}[1]{\tikz[baseline=(myanchor.base)] \node[circle,fill=.,inner sep=1pt] (myanchor) {\color{-.}\bfseries\footnotesize #1};}
\titlespacing{\section}{0pt}{2ex}{1ex}
\titlespacing{\subsection}{0pt}{0.5ex}{0.2ex}
\titlespacing{\subsubsection}{0pt}{0.5ex}{0ex}
\begin{document}

\title{ADC/DAC-Free Analog Acceleration of Deep Neural Networks with Frequency Transformation}
\author{Nastaran Darabi$^1$$^*$,~\IEEEmembership{Graduate Student Member,~IEEE,} Maeesha Binte Hashem $^1$$^*$,~\IEEEmembership{Graduate Student Member,~IEEE,} Hongyi Pan$^1$, Ahmet Cetin$^1$~\IEEEmembership{Fellow,~IEEE,} Wilfred Gomes$^2$, and Amit Ranjan Trivedi$^1$~\IEEEmembership{Senior Member,~IEEE}\\
$^1$ Department of Electrical and Computer Engineering, University of Illinois at Chicago, IL 60607 USA \\
$^2$ Intel, Santa Clara, CA 95054 USA
\thanks{$^*$ Both authors contributed equally to this work.
}}

\maketitle
\begin{abstract}
The edge processing of deep neural networks (DNNs) is becoming increasingly important due to its ability to extract valuable information directly at the data source to minimize latency and energy consumption. Although pruning techniques are commonly used to reduce model size for edge computing, they have certain limitations. Frequency-domain model compression, such as with the Walsh-Hadamard transform (WHT), has been identified as an efficient alternative. However, the benefits of frequency-domain processing are often offset by the increased multiply-accumulate (MAC) operations required. This paper proposes a novel approach to an energy-efficient acceleration of frequency-domain neural networks by utilizing analog-domain frequency-based tensor transformations. Our approach offers unique opportunities to enhance computational efficiency, resulting in several high-level advantages, including array micro-architecture with parallelism, ADC/DAC-free analog computations, and increased output sparsity. Our approach achieves more compact cells by eliminating the need for trainable parameters in the transformation matrix. Moreover, our novel array micro-architecture enables adaptive stitching of cells column-wise and row-wise, thereby facilitating perfect parallelism in computations. Additionally, our scheme enables ADC/DAC-free computations by training against highly quantized matrix-vector products, leveraging the parameter-free nature of matrix multiplications. Another crucial aspect of our design is its ability to handle signed-bit processing for frequency-based transformations. This leads to increased output sparsity and reduced digitization workload. On a 16$\times$16 crossbars, for 8-bit input processing, the proposed approach achieves the energy efficiency of 1602 tera operations per second per Watt (TOPS/W) without early termination strategy and 5311 TOPS/W with early termination strategy at VDD = 0.8 V.  
\end{abstract}

\section{Introduction}
In recent years, deep learning has gained significant traction in critical domains such as healthcare, finance, security, and autonomous vehicles \cite{chen2019deep,shukla2023robust,shukla2021ultralow,navardi2023mlae2,hassanalieragh2015health}. Especially as the complexity and accuracy requirements of deep learning applications continue to grow, deploying deep neural networks (DNNs) at the network's edge has become increasingly common for these applications. The edge, characterized by limited computing and storage resources, poses challenges for running large-scale DNN models efficiently. To overcome these limitations, pruning techniques have emerged as a popular approach to enhance edge computing for DNNs \cite{yu2020easiedge}. By selectively removing network parameters that contribute only minimally to prediction accuracy, pruning reduces the model's size and the computing/storage resources required for inference. 

Two main types of pruning techniques have been developed: unstructured and structured pruning \cite{anwar2017structured, liu2018rethinking}. Unstructured pruning removes connections with minimal magnitude weights, minimizing the impact on prediction accuracy. However, this technique does not always lead to proportional performance benefits, as it disrupts the organization of network weights, affecting their mappability to modular computing substrates like GPUs, FPGAs, or spatial digital processors. On the other hand, structured pruning removes entire channels, filters, rows, or columns in fully connected layers, thereby preserving the data organization structure for mappability \cite{han2015deep}. However, structured pruning is susceptible to over-pruning. The constraint of removing entire structures, such as weight channels, may inadvertently eliminate salient weights and connections, creating a trade-off between model size reduction and accuracy. This limitation becomes particularly significant in applications like object detection or segmentation, where fine-grained representations are crucial for achieving high performance \cite{tanaka2020pruning, tayebati2023hybrid, wen2016learning, carreira2018learning}.
\begin{figure}[t!]
    \centering
    \includegraphics[width=\linewidth]{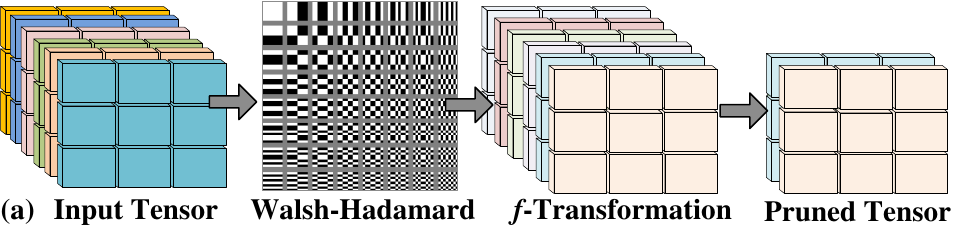}
    \includegraphics[width=0.49\linewidth]{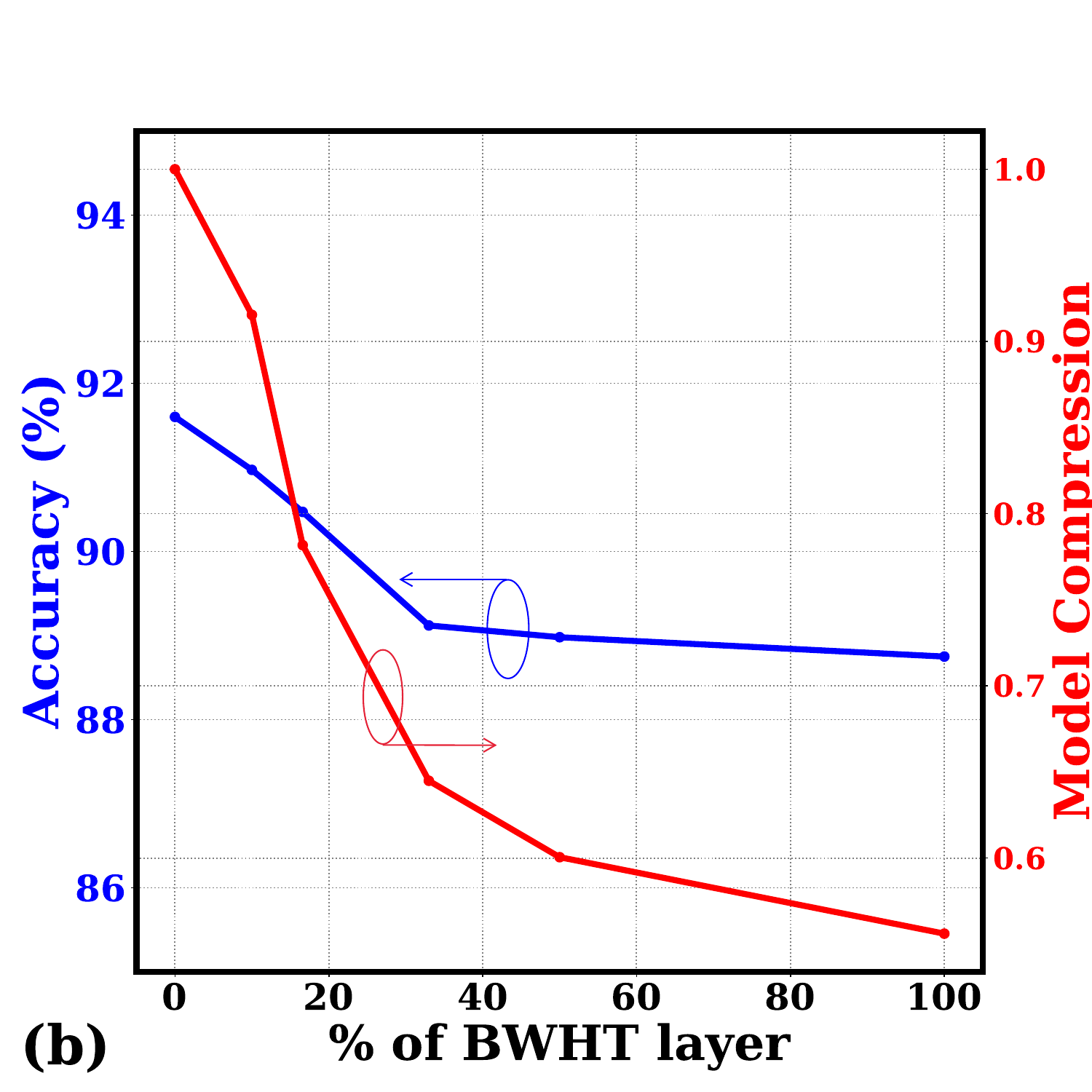}
    \includegraphics[width=0.49\linewidth]{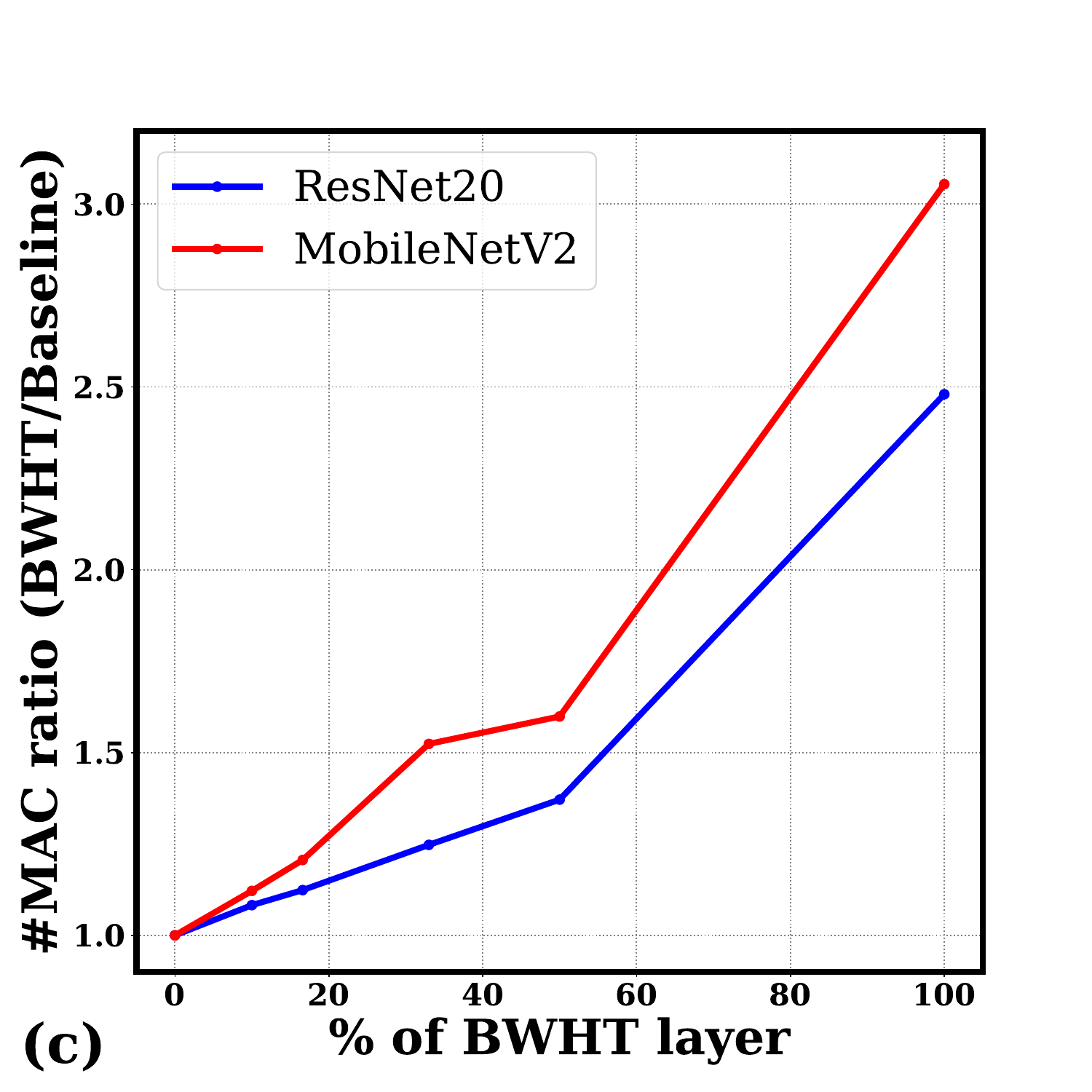}
    \caption{\textbf{Frequency-domain processing of neural networks:} \textbf{(a)} Frequency transformations of neural tensors. \textbf{(b)} Prediction accuracy and model compression by increasingly processing more layers of ResNet20 with Walsh-Hadamard transforms (WHT). The accuracy is characterized for the CIFAR10 dataset. Model compression is the ratio of the number of parameters in the frequency domain processing over conventional processing. \textbf{(c)} Increase in multiply-accumulate (MAC) operations under frequency domain processing compared to conventional processing for MobileNetV2 and ResNet20 by increasingly processing more layers in the frequency domain.}
    \label{fig:Introduction}
\end{figure}

Frequency-domain model compression has emerged as an efficient alternative to traditional model pruning techniques \cite{chen2016compressing, liu2018frequency, xu2020learning}. Frequency-based compression leverages fast algorithms for transforms such as the discrete cosine transform (DCT) or discrete Fourier transform (DFT), effectively identifying and removing redundant or uncorrelated information in the frequency domain. Various frequency transformations have been applied to DNNs, including Walsh-Hadamard transform (WHT) \cite{rossi2020walsh, pan2021fast}, discrete cosine transform (DCT) \cite{pan2022dct, koizumi2018end}, block circulant matrices (BCM) \cite{8203813}, discrete Fourier transform (DFT) \cite{xu2019frequency}, discrete wavelet transform (DWT) \cite{mohsen2018classification}, singular value decomposition (SVD) \cite{xue2013restructuring}, and principal component analysis (PCA) \cite{riera2022dnn}. 

For instance, Fig. \ref{fig:Introduction}(b) illustrates the potential benefits of implementing the layers of ResNet20 to the transform domain using Walsh-Hadamard transforms (WHT), which approximately computes the frequency domain parameters and convolutions can be approximately computed in the WHT domain as element-wise multiplications. By progressively processing more layers in the frequency domain, a remarkable reduction in the model size can be achieved while only incurring a limited loss in accuracy. In the figure, when all network layers are processed in the frequency domain, the number of trainable parameters reduces by 55.6\% while incurring only a $\sim$3\% accuracy loss for the classification of the CIFAR10 dataset. The figure also demonstrates that frequency domain processing of a typical neural network can be applied on a \textit{curve}. Thereby, under accuracy constraints, an optimal number of layers can be processed in the frequency domain to achieve maximal model compression. Moreover, WHT, as considered in this example, is well-suited for low-power and computationally efficient processing, as the transformation matrices only consist of binary values. More detailed characterization of WHT-based frequency transformation was presented in \cite{pan2022block,pan2023hybrid}. Motivated by these findings, this work primarily focuses on WHT-based model compression.

While frequency-domain model compression using WHT provides substantial advantages in model compression, it also introduces a notable increase in the required multiply-accumulate (MAC) operations. Fig. \ref{fig:Introduction}(c) compares MAC operations with conventional processing and frequency domain processing when applying frequency transforms to the bottleneck layers of MobileNetV2 and the residual layers of ResNet20. On average, the MAC operations increase three-fold under the transforms for MobileNetV2 when all layers are processed in the frequency domain. This significant increase in computational workload offsets the benefits of frequency domain model size reduction, emphasizing the critical need for computationally efficient transformations to leverage this technique fully. 

We present a novel analog acceleration approach to address this challenge and make the following key contributions:
\begin{itemize}[noitemsep,topsep=0pt,leftmargin=10pt]
    \item \textit{Analog acceleration of frequency-domain tensor transformation:} We leverage the analog representation of operands to simplify the frequency transformation of neural tensors. E.g., by leveraging Kirchhoff's law, we obviate overheads of a digital multiplier to use a transistor for analog-domain multiplications between operands. Similarly, weight-input product terms represented as charges in our scheme are summed over a wire without a dedicated adder. Thus, exploiting physics minimizes the workload, processing elements, and data movements.
    \item \textit{Matrix-level operational parallelism:} Analog processing for operands in our scheme also simplifies the design of computing cells, enabling a large crossbar to operate within a limited area and energy budget. Additionally, our novel crossbar micro-architecture facilitates matrix-level operational parallelism through adaptive stitching of computing cells along both column and row dimensions. This enables parallel processing on all input vector elements and parallel determination of all output vector elements.
    \item \textit{Co-designing of analog crossbar operations and compute flow:} Even though internal operands are represented in the analog domain, the crossbar computations are performed without the need for ADC/DAC (Analog-to-Digital and Digital-to-Analog) conversions using a co-design approach. This eliminates the overhead of domain conversion and allows for better technology node scalability. We adopt input bitplane-wise processing directly on digital input vector streams to achieve DAC-free computations. Our computations are ADC-free by training against extremely quantized matrix-vector products and leveraging matrix multiplications' parameter-free nature in frequency domain neural processing. Our crossbar design supports signed-bit processing, which promotes increased output sparsity and reduces the workload. Additionally, we present a novel training loss function that promotes output sparsity, further capitalizing on these co-design aspects. 
\end{itemize}

Sec. II presents the necessary background for the proposed techniques. Sec. III discusses the architecture for analog-domain frequency transforms. Sec. IV presents simulation results. Sec. V concludes.

\begin{figure}[t!]
    \centering
    \includegraphics[width=\linewidth]{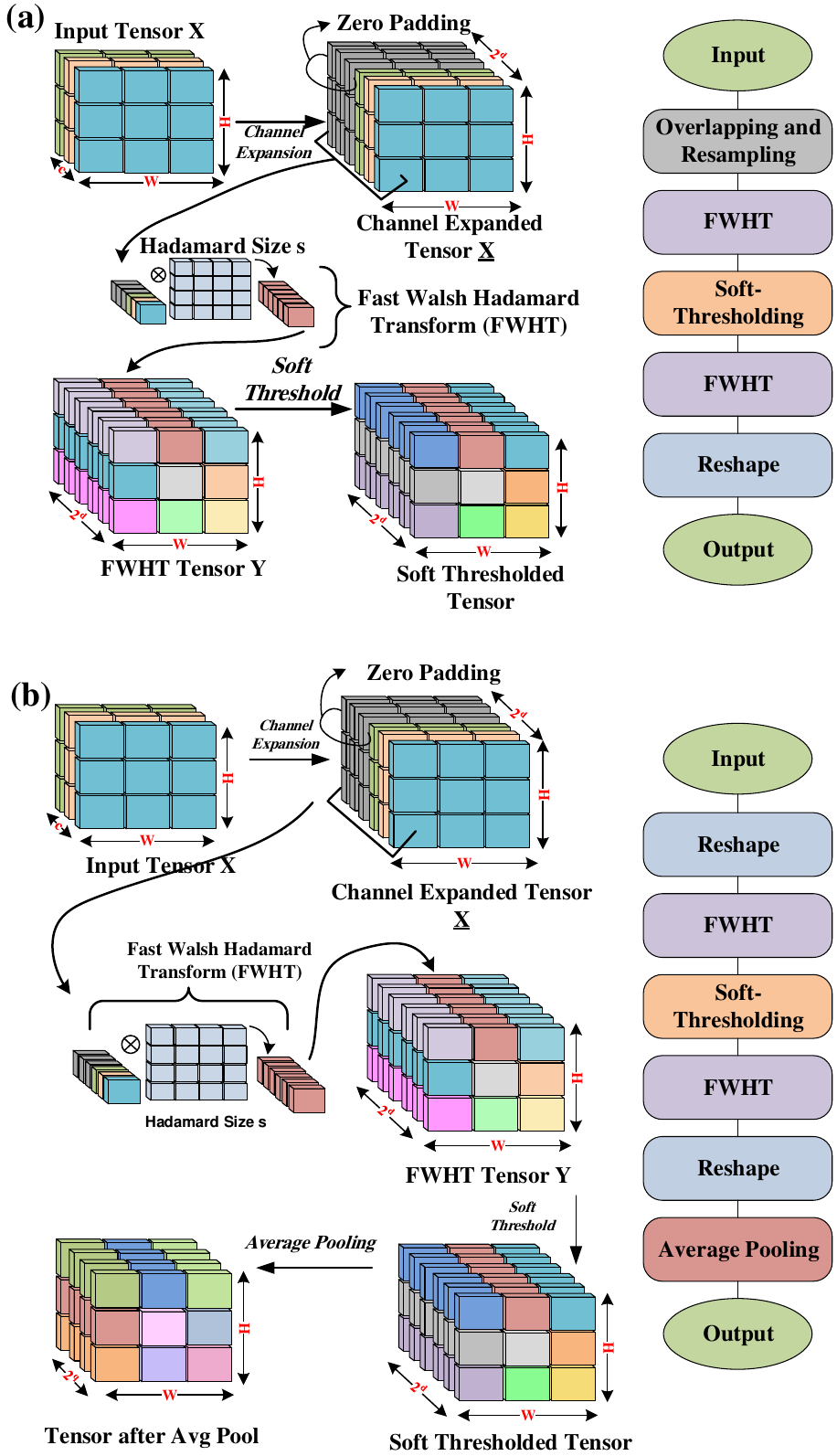}
    \caption{ \textbf{(a)} Channel expansion and \textbf{(b)} channel projection under one-dimensional Blockwise Hadamard Transformation (BWHT). The figure pictorially shows the transformation of input tensors under expansion and projection steps by zero-padding, Hadamard multiplications, and soft thresholding. A flow diagram for tensor transformations is shown to the right.}
    \label{fig:Algorithm}
\end{figure}

\begin{figure}[h!]
    \centering
    \includegraphics[scale=0.5]{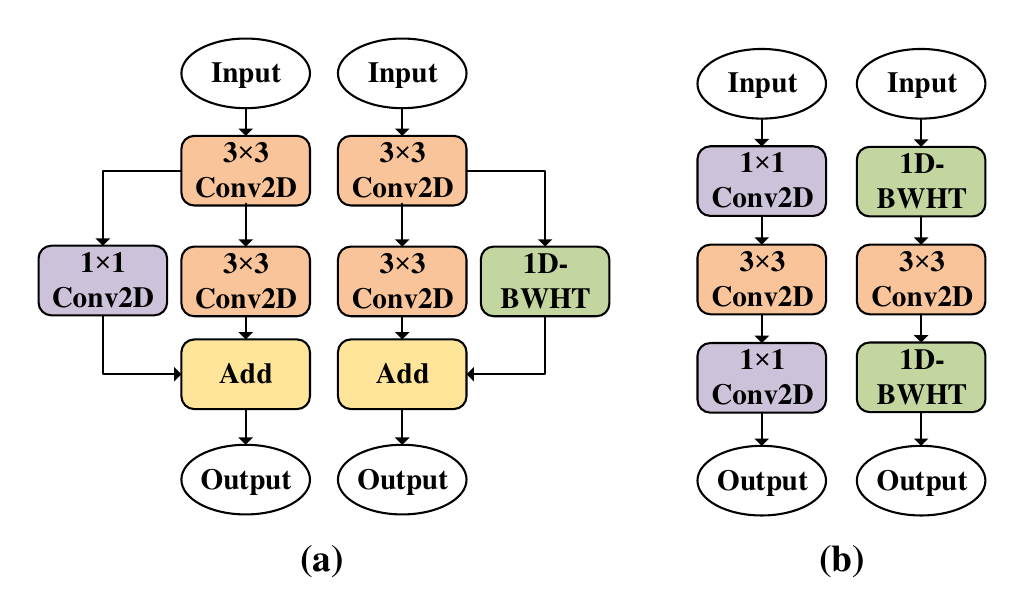}
    \caption{ \textbf{(a)} Residual block of ResNet20 with $1\times1$ convolution layers, in our design $1\times1$ convolution layers are replaced with 1D-BWHT layer in ResNet20. \textbf{(b)} MobileNetV2 bottleneck layers \textit{vs.} MobileNetV2 bottleneck layers with 1D-BWHT layer. BWHT can replace $1 \times 1$ convolution layers, achieving similar accuracy with fewer parameters.}
    \label{fig: MobileNet layers}
\end{figure}

\begin{figure*}[t!]
    \centering
    \includegraphics[width=\linewidth]{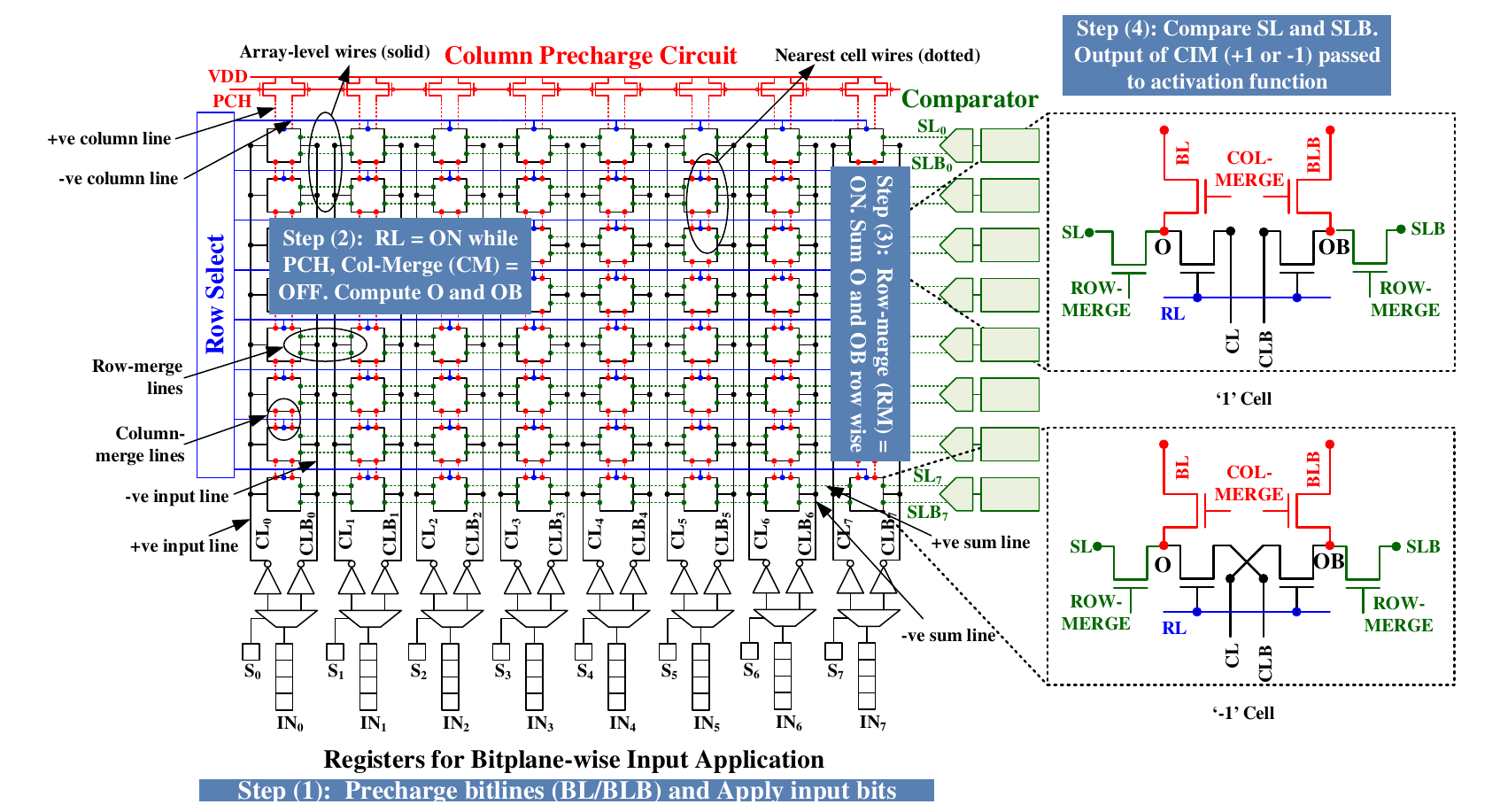}  
    \caption{\textbf{Architecture and operation flow for an analog acceleration of frequency-domain neural processing:} The operation consists of four steps: (1) precharging bit lines (BL/BLB) and applying input, (2) enabling parallel local computations in O and OB, (3) activating row-merge to connect all cells row-wise and summing O/OB in sum lines (SL/SLB), and (4) comparing SL/SLB values and applying soft thresholding for accurate output generation.}
    \label{fig:top-level architecture}
\end{figure*}

\section{Background}
\subsection{Walsh-Hadamard Transform (WHT)}
WHT shares similarities with the Fast Fourier Transform (FFT) in that both can convert convolutions in the time or spatial domain into multiplications in the frequency domain. However, a major advantage of the WHT over other transformations is that its transform matrix comprises exclusively binary values (-1 and 1). This makes the transformations essentially multiplication-free, resulting in higher efficiency. 

Let $X, Y \in \mathbf{R}^m$ be the vector in the time-domain and WHT-domain, respectively, where \textit{m} should be an integer power of 2 ($2^k, k\in \mathbf{N}$). Under WHT: 
\begin{equation}
Y = W_k X
\end{equation}
Here $W_k$ is a $2^k \times 2^k$ Walsh matrix. For WHT of $X$, a Hadamard matrix $H_k$ is constructed using the following procedure:
\begin{equation}
H_k = \left\{
        \begin{array}{ll}
            1, & k=0\\
           \begin{bmatrix}
            $$H_{k-1}$$ & $$H_{k-1}$$\\
            $$H_{k-1}$$ & $$-H_{k-1}$$
        \end{bmatrix}, & k>0
        \end{array}
    \right.
\end{equation}
Here, we rearrange matrix rows to increase the sign change order, resulting in the Walsh matrix derived from the Hadamard matrix as generated in Eq. (2). The process entails partitioning the Hadamard matrix of size $N$ into four sub-matrices of dimensions $N/2 \times N/2$, followed by replacing the upper-left and lower-right sub-matrices with their negative counterparts, i.e., negating each entry. These steps are repeated until each sub-matrix is reduced to a size of $1 \times 1$. The resulting matrix exhibits the unique property that every row of the matrix is orthogonal to each other, with the dot product of any two rows being zero. This orthogonality renders the Walsh matrix particularly advantageous in a wide range of signal and image processing applications \cite{jayathilake2013discrete}.

WHT presents a computational challenge when the dimension of the input vector is not a power of two. A technique called blockwise Walsh-Hadamard transforms (BWHT) was introduced to address this issue in \cite{pan2022block}. The BWHT approach divides the transform matrix into multiple blocks, each sized to an integer power of two. By partitioning the transform matrix, only the last block requires zero padding if the input vector's dimension is not a power of two. This partitioning strategy significantly reduces the worst-case size of operating tensors, mitigating excessive zero-padding. 

\subsection{Frequency-Domain Compression of Deep Neural Networks}
The BWHT-based frequency transformations in Fig. \ref{fig:Algorithm} can expand or project channels. The expansion operation typically uses a 1D-BWHT layer to increase the number of channels in the feature map, providing a richer representation for subsequent layers to learn from. On the other hand, the projection operation employs a 1D-BWHT layer to reduce the dimensional to make the network computationally efficient while retaining essential features. In Pan et al.'s study \cite{pan2022block}, these transformations maintained a matching accuracy under frequency transforms while achieving significant compression than standard implementation on benchmark datasets such as CIFAR-10, CIFAR-100, and ImageNet. The number of parameters in the BWHT layer is thus proportional to the thresholding parameter $T$, which is significantly smaller than the number of parameters in a $1 \times 1$ convolution layer. The activation function $S_T$ for this frequency-domain compression is given as
\begin{equation}
y = S_{T}(x) = sign(x)(|x| - T) = \left\{
                                        \begin{array}{ll}
                                            x+T, & x < -T\\
                                            0, & |x| \le T\\
                                            x-T, & x > T
                                        \end{array}
                                    \right.
\end{equation}

Frequency-domain transformations, such as BWHT, can be integrated into state-of-the-art deep learning architectures to compress them. For example, in the context of MobileNetV2, which uses $1 \times 1$ convolutions in its bottleneck layers to reduce computational complexity, BWHT can replace these convolution layers while achieving similar accuracy with fewer parameters, as shown in Fig. \ref{fig: MobileNet layers}(b). Unlike $1 \times 1$ convolution layers, which have a parameter count proportional to the product of input feature map size and the number of output channels, BWHT-based binary layers use fixed Walsh-Hadamard matrices, eliminating trainable parameters. Instead, a soft-thresholding activation function with a trainable parameter $T$ can be used to attend to important frequency bands selectively. We opt for soft-thresholding over the ReLU activation function because the magnitude of the coefficients in the transform domain holds significant importance. This approach allows us to preserve high-amplitude negative coefficients in the transform domain, which are crucial for our analysis. Similarly, in ResNet20, we can replace $1 \times 1$ convolutions with BWHT layers. This modification is illustrated in Fig. \ref{fig: MobileNet layers}(a).  

However, BWHT transforms also increase the necessary computations for deep networks, as demonstrated in Fig. \ref{fig:Introduction}(c). In Sec. III, we discuss how micro-architectures and circuits can enhance the computational efficiency of BWHT-based tensor transformations.

\section{Analog-Domain Frequency Transforms}
Analog domain processing has emerged as a promising approach to accelerate vector and matrix-level parallel instructions suitable for low-precision computations, such as matrix-vector products in deep neural networks (DNNs). For instance, analog computations simplify processing cells and enable the integration of storage and computations within a single cell in many compute-in-memory designs. This significantly reduces data movement during deep learning computations, which is a critical bottleneck for energy and performance in traditional processors \cite{9987520, 9833492,10168632}.

 
However, a major drawback of analog domain processing is its dependence on analog-to-digital converters (ADCs) and digital-to-analog converters (DACs) for domain conversions of operands. These conversions are necessary when interfacing with digital systems or processing digital data, as analog signals must be converted to a digital format for processing and vice versa. Nevertheless, ADC and DAC operations introduce design complexities, significant area/power overheads, and limitations in technology scalability. Moreover, the performance of ADCs and DACs is constrained by speed, power consumption, and cost, further limiting the overall capabilities of analog domain computations.
\begin{figure}[t!]
\centering
    \includegraphics[width=0.9\linewidth]{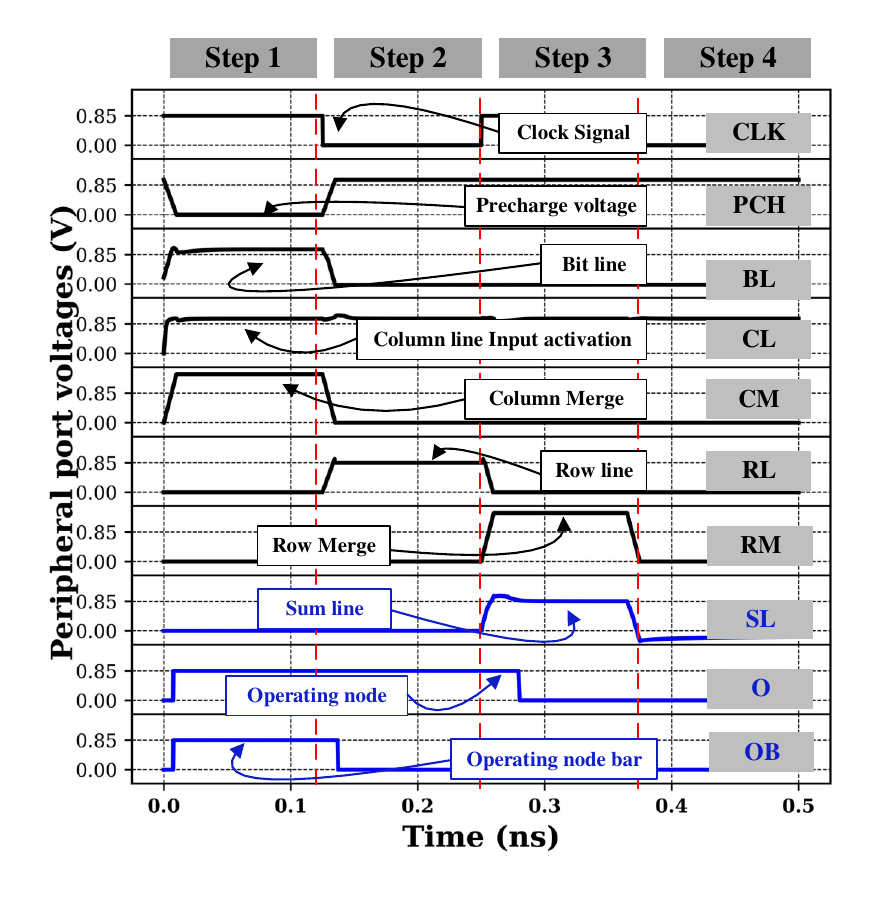}   
    \caption{\textbf{Timing diagram of signal flows:} Waveforms of key signals in the CIM operation, including clock signal (CLK), precharge signal (PCH), bit lines (BL/BLB), column lines (CL/CLB), column-merge signal (CM), row lines (RL), row-merge signal (RM), sum lines (SL/SLB), and operating points (O/OB). The four-step CIM operation is completed in two clock cycles. The compact cell design is comprised solely of NMOS transistors. }
    \label{fig: waveforms}
\end{figure}

In the following discussion, we present our proposed techniques for analog domain processing of frequency operations, eliminating the need for ADC/DAC conversions, even when operating on digital input vectors and computing output vectors in the digital domain. To achieve this, our approach incorporates bitplane-wise input vector processing and co-designs learning methods that can operate accurately under extreme quantization, enabling ADC/DAC-free operations.

\subsection{Crossbar Micro-architecture Design and Operation}
In Fig. \ref{fig:top-level architecture}, we leverage analog computations for frequency domain processing of neural networks. The crossbar in the design combines six transistors (6T) NMOS-based cells for analog-domain frequency transformation. The corresponding cells for `-1' and `1' entries in the Walsh-Hadamard transform matrix are shown to the figure's right. The crossbar combines these cells according to the elements in the transform matrix. Since the transform matrix is parameter-free, computing cells in the proposed design are simpler by being based only on NMOS for a lower area than conventional 6T or 8T SRAM-based compute-in-memory designs. Additionally, processing cells in the proposed crossbar are \textit{stitchable} along rows and columns to enable perfect parallelism and extreme throughput. 

Crossbar's operation comprises four steps. These steps are marked in Fig. \ref{fig:top-level architecture}. In the \textit{first step}, the precharge signal (PCH) is activated, charging all the bit lines (BL and BLB) to the supply voltage (VDD). The column merge signal (CM) is set to high, effectively connecting or ``stitching'' all cells along a column. The input vector is loaded bitwise along the column peripherals. Depending on the sign bit of each element in the input vector, CL or CLB is applied with the corresponding magnitude bit, CL is activated for positive input vector elements, and CLB is activated for negative elements. In the \textit{second step}, the row line (RL) is set to high while the CM and PCH signals are turned off, thus disconnecting cells along a column to allow independent local computations to take place in parallel. Depending on CL/CLB, cell output nodes O and OB either retain the precharge voltage or are grounded. In the \textit{third step}, cell outputs O and OB are summed row-wise by activating the row merge signal (RM) while turning OFF CM and RL. The potential of O and OB are then averaged row-wise on sum lines, SL and SLB, by stitching the computing cells row-wise using a row-merge signal in the figure. In the \textit{fourth and final step}, the values at SL and SLB are compared using a single comparator, resulting in single-bit output. 

Fig. \ref{fig: waveforms} shows the signal flow diagram for the above four steps. The 16 nm predictive technology models (PTM) simulation results are shown using the low standby power (LSTP) library in \cite{PTM} and VDD = 0.85V while boosting RM and CM signals. Unlike comparable compute-in-memory designs such as \cite{9972346}, which place product computations on bit lines, in our design, these computations are placed on local nodes in parallel at all array cells. This improves parallelism, energy efficiency, and performance by computing on significantly less capacitive local nodes than bit lines of traditional designs.     

\begin{figure}[t!]
    \centering
    \includegraphics[width=0.9\linewidth]{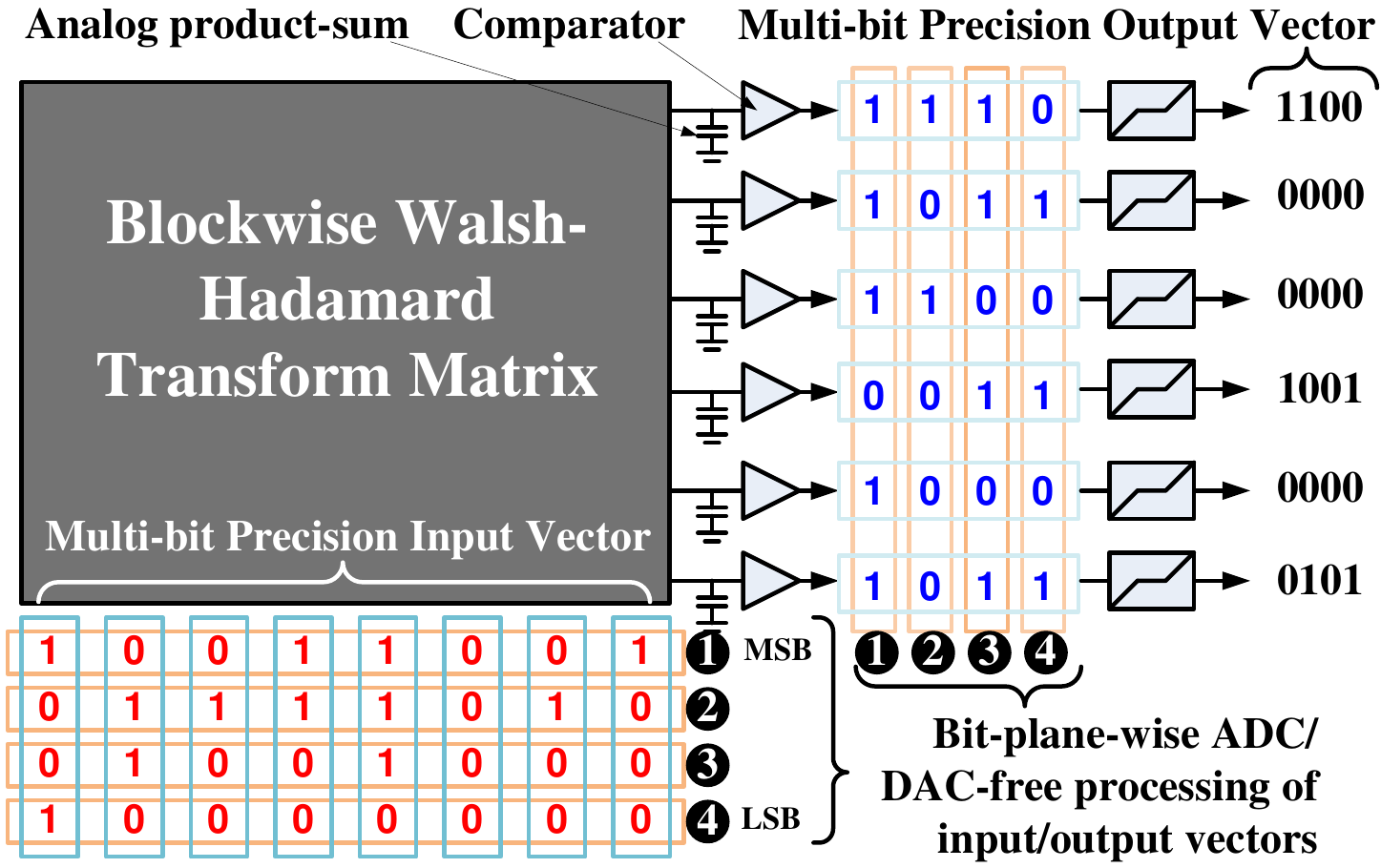}
    \caption{\textbf{Bitplane-wise operation flow:} A multi-bit input vector is processed in a bitplane-wise manner. The same significant bits of vector elements are grouped together and processed in a single step. Analog the crossbar rows output vector bits are computed in parallel. The output bits generated for all input bitplanes are concatenated to produce the multi-bit output vector.}
    \label{fig:operation-flow}
\end{figure}

\subsection{ADC-Free by Training against 1-bit Quantization}
Fig. \ref{fig:operation-flow} presents the high-level operation flow of our scheme for processing multi-bit digital input vectors and generating the corresponding multi-bit digital output vector using the analog crossbar illustrated in Fig. \ref{fig:top-level architecture}, without the need for ADC and DAC operations. The scheme utilizes bitplane-wise processing of the multi-bit digital input vector and is trained to operate effectively with extreme quantization. In the figure, the input vector's elements with the same significance bits are grouped and processed in a single step using the scheme described in Fig. \ref{fig:top-level architecture}, which spans two clock cycles. The analog charge-represented output is computed along the row-wise charge sum lines and thresholded to generate the corresponding digital bits. This extreme quantization approach is applied to the computed MAC output, eliminating the need for ADCs. With multiple input bitplanes, labeled as $\ballnumber{1}-\ballnumber{2}-\ballnumber{3}-\ballnumber{4}$ in the figure, the corresponding output bitplanes are concatenated to form the final multi-bit output vector.

\begin{figure}[t!]
    \centering  
    \includegraphics[width=0.49\linewidth]{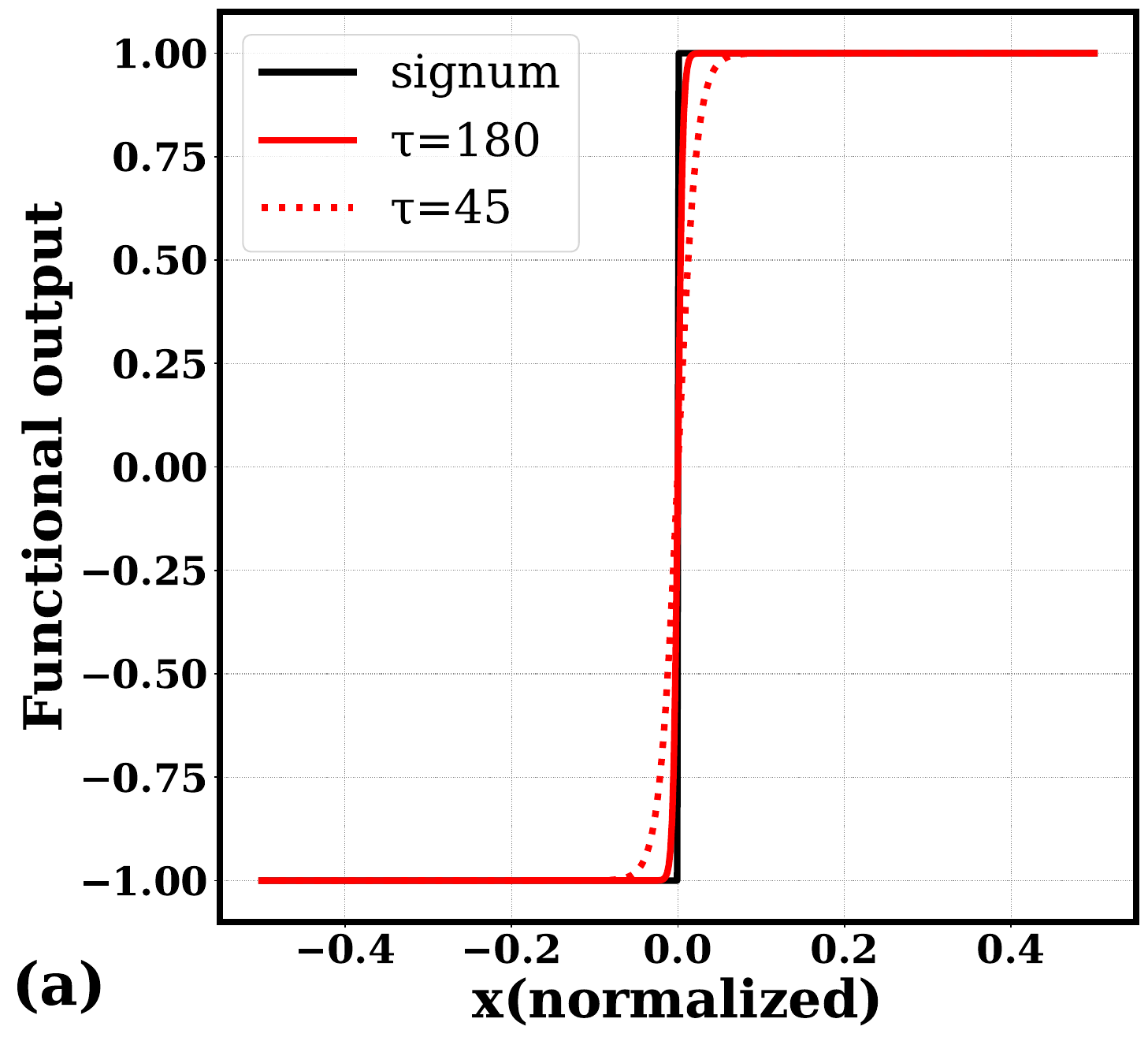}
    \includegraphics[width=0.45\linewidth]{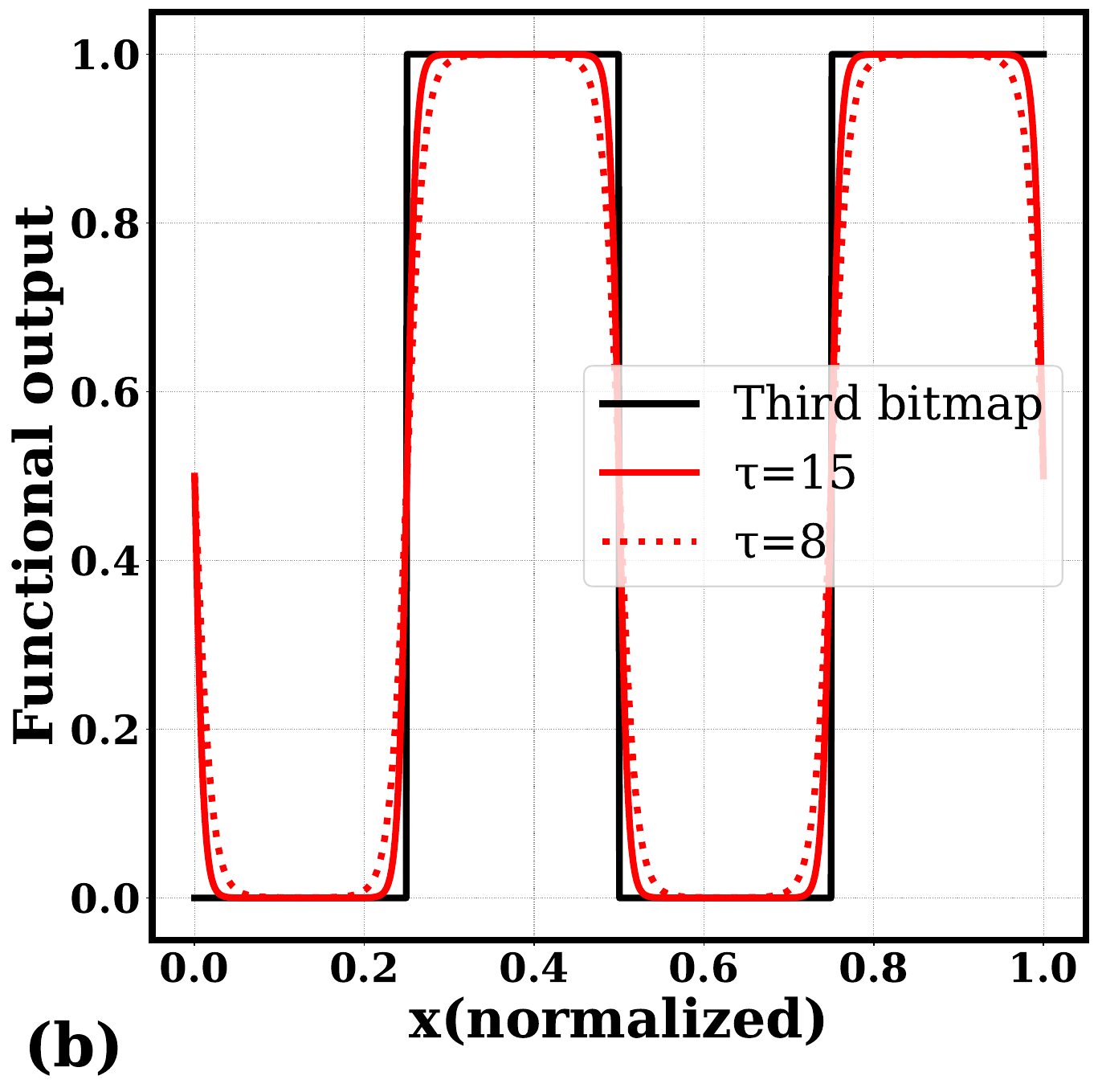}  
    \caption{\textbf{Approximation functions:} Continuous function approximation to \textbf{(a)} sign and \textbf{(b)} quantization function with well-defined derivatives to accelerate loss convergence.}
    \label{fig:approx_functions}
\end{figure}

Due to the extreme quantization applied to the analog output in the above scheme, the resulting digital output vector only approximates the true output vector under transformation. However, unlike a standard deep neural network (DNN) weight matrix, the transformation matrix used in our frequency domain processing is parameter-free. This characteristic enables training the system to effectively mitigate the impact of the approximation while allowing for a significantly simpler implementation without needing ADCs or DACs. The following training methodology achieves this:

Consider the frequency-domain processing of an input vector $\vectorbold{x_i}$. As shown in Fig. 2, we process $\vectorbold{x_i}$ by transforming it to the frequency domain, followed by parameterized thresholding, and then reverting the output to the spatial domain. Consider a DNN with $n$ layers that chain the above sequence of operations as $\vectorbold{x_{i+1}} = F_0(S_{T,i}(F_0(\vectorbold{x_{i}}))) $. Here, $F_0()$ is a parameter-free \textit{approximate} frequency transformation as followed in our scheme in Fig. 6. $S_{T,i}()$ is a parameterized thresholding function at the i$^{th}$ layer whose parameters $T_i$ are learned from the training data. 

Following the stochastic gradient descent (SGD)-based training frameworks to minimize the loss function $\mathcal{L}(\vectorbold{x_n})$ involves the derivatives of $S_{T}()$ and $F_0()$. Since $S_{T}()$ is a smooth activation function, its derivatives are well-defined. However, due to a bitplane-wise processing of quantized $\vectorbold{x_n}$ and 1-bit quantization of the resultant dot products, $F_0()$ is a non-continuous function with ill-defined derivatives. Therefore, we can approximate the partial derivatives of $F_0()$ to accelerate the loss convergence. 

\begin{figure}[t!]
    \centering  
    \includegraphics[width=0.47\linewidth]{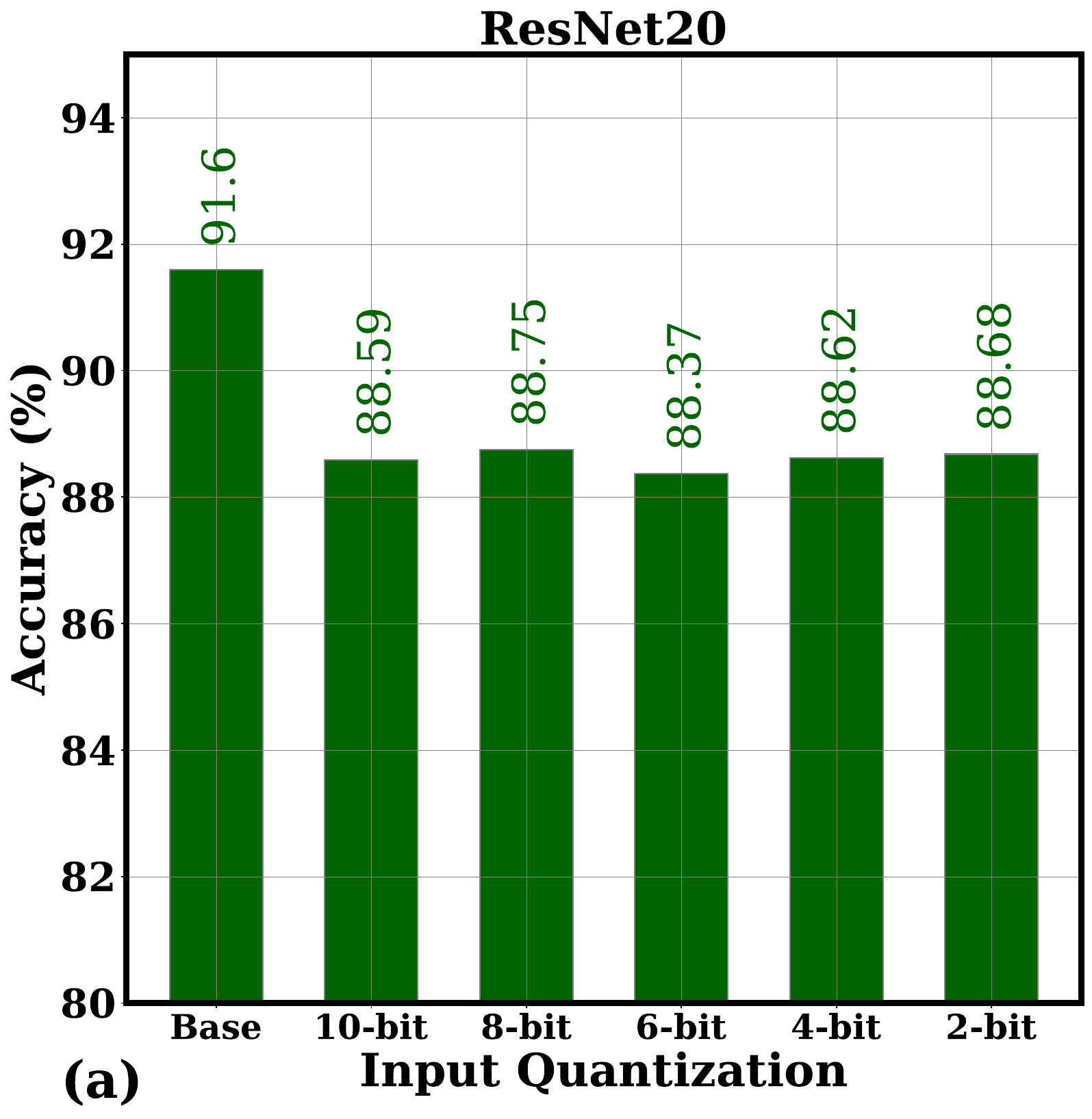}
    \includegraphics[width=0.49\linewidth]{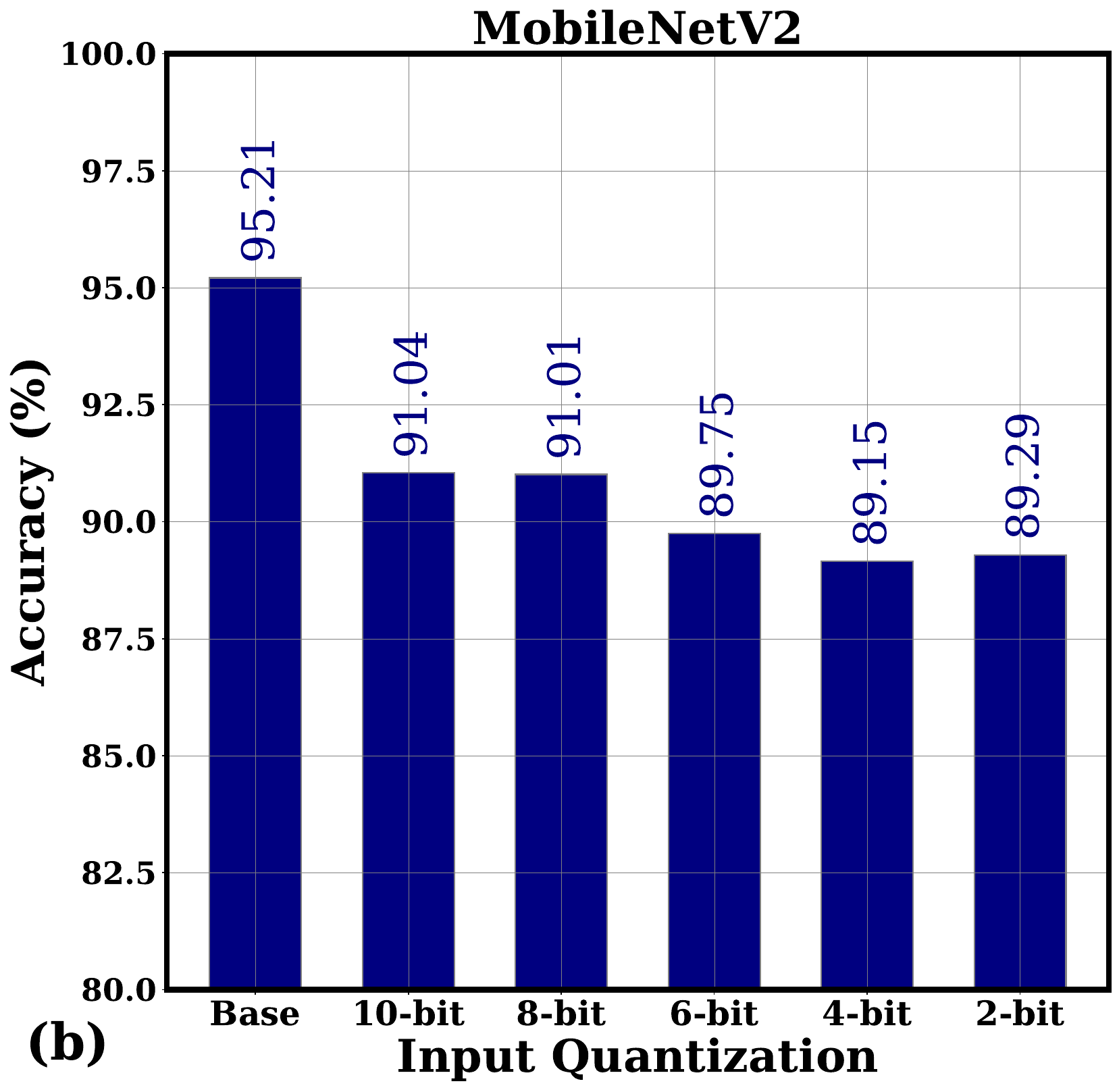}  
    \caption{\textbf{Accuracy under training with 1-bit quantization:} Impact of input quantization on the performance of deep learning models \textbf{(a)} ResNet20 and \textbf{(b)} MobileNetV2 on the CIFAR-10 dataset, while considering 1-bit product-sum quantization and varying input quantization levels. The results demonstrate that accuracy converges to a similar level across all input quantization levels, and it is 3-4\% lower than the floating-point baseline. }
    \label{fig:Accuracy results for MobileNet and ResNet}
\end{figure}

\begin{figure*}[h!]
    \centering
    \includegraphics[height=0.3\linewidth]{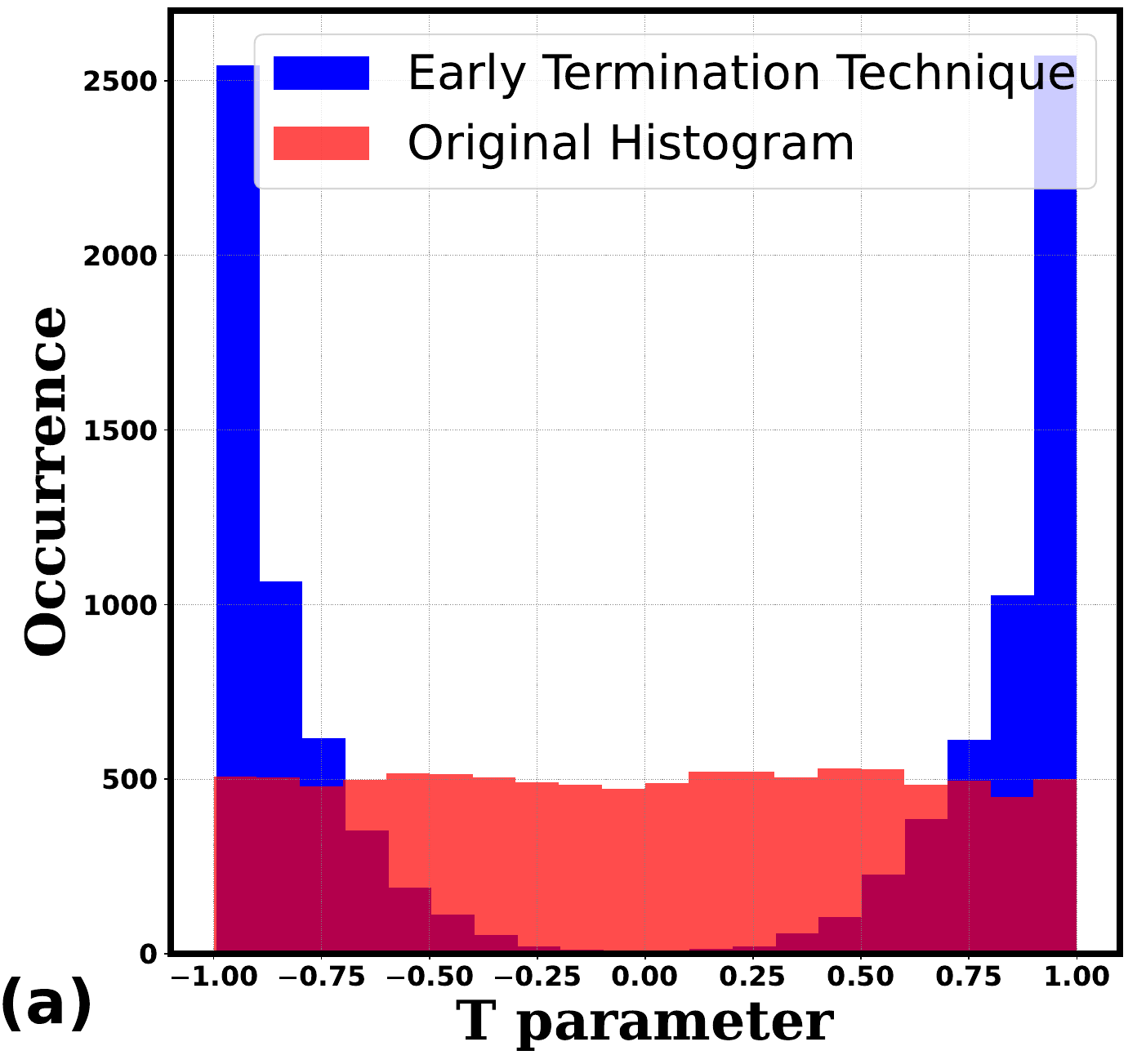}
    \includegraphics[height=0.3\linewidth]{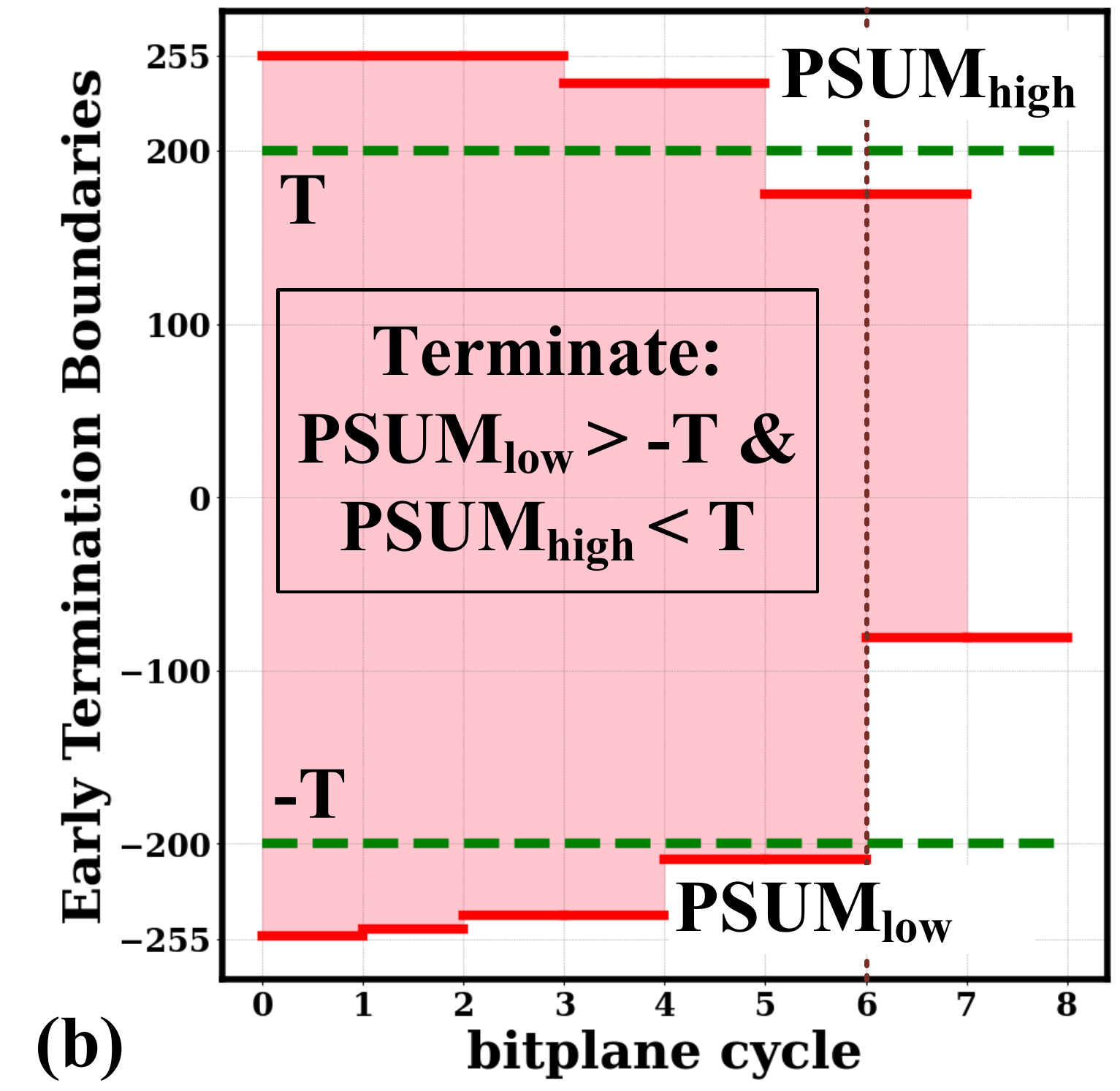}
    \includegraphics[height=0.3\linewidth]{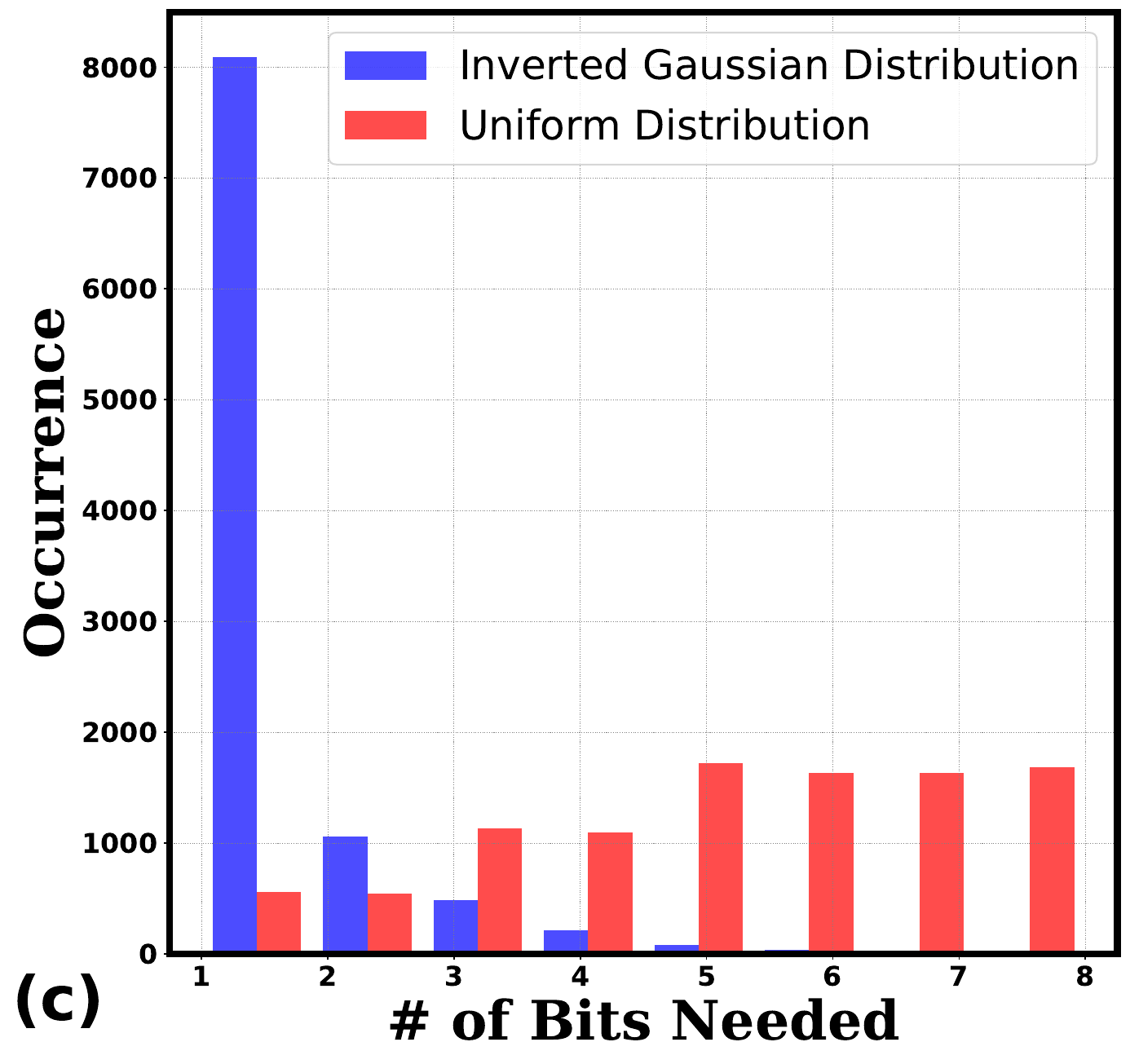}
    \caption{\textbf{Predictive Early Termination:} \textbf{(a)} Distribution of thresholding parameters with and without early termination-based loss function. A modified loss function pushes $T$-values towards 1 or -1, aiding the sparsity or neuron output. \textbf{(b)} Comparison bounds determining the early termination with varying bitplane cycles. As the processing moves to lower significance bitplanes, progressively, the comparison bounds tighten, thus increasing the opportunities for early termination. \textbf{(c)} Histogram illustrating the number of bits required for early termination using 10,000 randomly generated 8-bit inputs and weights. The analysis was conducted under two scenarios: 1) when the soft thresholding parameter (T) follows a uniform distribution and 2) when it adheres to an Inverted Gaussian distribution, as dictated by our loss function.}
    \label{fig: Early Termination Process}
\end{figure*}

For input $\vectorbold{x}$, the i$^{th}$ element of the output vector $F_0(\vectorbold{x})$ under our approximate frequency transform can be defined as
\begin{equation}
    F_{0,i}(\vectorbold{x}) = \sum_{b=1}^B \text{sign}
    \Bigg(\sum_{j=1}^N I_{jb} B_{ij}\Bigg) \times 2^{b-1}
\end{equation}
Here, $j$ indices over BWHT matrix columns running up to $N$, and $b$ indices over input bits running up to $B$. $I_{jb}$ is the b$^{th}$ bit of digitized input vector $\vectorbold{x}$ at column $j$. $B_{ij}$ is BWHT matrix entry at i$^{th}$ row and j$^{th}$ column. sign() function's output is one if the operand is positive; otherwise, the output is -1. Now consider the partial derivative of $F_{0,i}()$ w.r.t. j$^{th}$ element of $\vectorbold{x}$, i.e., $x_{j}$:    
\begin{subequations}
\begin{align}
\frac{\partial F_{0,i}}{\partial x_{j}} &= \frac{\partial}{\partial x_{j}}\sum_{b=1}^B \text{sign}\Bigg(\sum_{j=1}^N I_{jb} B_{ij}\Bigg) \times 2^{b-1}\\
    &=\sum_{b=1}^B \text{sign}^\prime
    \Bigg(\sum_{j=1}^N I_{jb} B_{ij}\Bigg) \times \frac{\partial I_{jb}}{\partial x_{j}} \times 2^{b-1}
\end{align}
\end{subequations}
As can be noticed that the partial derivatives involve discontinuous functions sign() and quantization functions. To handle these discontinuities, we approximated these functions with continuous functions as follows:  
\begin{equation}
sign(x) = \lim_{{\tau} \to \infty} tanh(x \cdot \tau)
\end{equation}
\begin{equation}
I_b(x) = \lim_{{\tau} \to \infty} \frac{exp(-\tau \cdot sin(2\pi \cdot 2^{b_{max}-b} \cdot x/ x_{max}))}{1+exp(-\tau \cdot sin(2\pi \cdot 2^{b_{max}-b} \cdot x/ x_{max}))}
\end{equation}
Here, the sign() function is approximated as tanh(). Quantization functions corresponding to various significance bits are approximated as in Eq. (7). The above approximation functions involve hyperparameter $\tau$, which matches the involved functions accurately when it tends to infinity. For loss function minimization over several training iterations, $\tau$ can be incrementally increased to avoid creating sharp local minima. Fig. \ref{fig:approx_functions} plots the approximating function. For $I_b()$, results are shown for the quantization function corresponding to the second most significant bit. Fig. \ref{fig:Accuracy results for MobileNet and ResNet} shows the accuracy under training with 1-bit quantization of product-sum for ResNet20 and MobileNetV2 on the CIFAR-10 dataset. The results demonstrate that accuracy converges to a similar level across all input quantization levels, and it is 3-4\% lower than the floating-point baseline.

\subsection{Predictive Early Termination by exploiting Output Sparsity}
In Fig. 6, the BWH-transformed output ($x$) undergoes filtering via an activation function $S_T()$. This activation function yields a zero output when $|x| \le T$. Consequently, we can anticipate a substantial level of output sparsity within the frequency domain processing under consideration. To leverage this sparsity, in the following, we discuss an early termination strategy that enables us to avoid processing all input bitplanes. Instead, we can predictably terminate the processing when we anticipate a zero output, decreasing computation time and energy consumption. 

For the predictive early termination scheme, the execution begins with processing the input vector's most significant bitplane (MSB) and progresses to increasingly lower-significance bitplanes. At the currently executed bitplane $b$, the running output of frequency processing is computed as $y_b = \sum_{b=b}^B O_b \times 2^{b-1}$. Here, $O_b \in $ [-1, 1] is the binary output computed during the $b^{th}$ input bitplane processing by thresholding the computed analog sum on the sum lines. Based on the current running sum output, the expected upper and lower bounds of $y_b$ are computed as $y_{b,UB} = \sum_{k=b}^B O_k \times 2^{k-1} + \sum_{k=0}^{b-1} 2^{k-1}$ and $y_{b,LB} = \sum_{k=b}^B O_k \times 2^{k-1} - \sum_{k=0}^{b-1} 2^{k-1}$. Next, the bounds on the running sum are compared to the respective thresholding parameter $T$. If $y_{b,UB} \le T$ and $y_{b,LB} \ge -T$, processing of the corresponding output element under frequency transformation can be terminated early since it is expected to be zero post-activation. 

Moreover, adjusting thresholding parameters can further improve the opportunities for early termination. Specifically, if the thresholding parameters ($T$) of $S_T()$ can be maximized during training, the output sparsity and early termination opportunities are maximized. A $T$-dependent regularizer term is added to the loss function for this,
\begin{equation}
    \mathcal{L}_{mod} = \mathcal{L}_{acc}(T)  -  \lambda \log \left( \sqrt{\frac{1}{g(T)^3}} \exp \left( -\frac{g(T)}{2} \right)\right).
\end{equation}
Here, $\mathcal{L}_{acc}$ is accuracy-dependent loss based on cross-entropy for classification. Through the second term, $T$ values tend to gravitate towards either 1 or -1 during training to follow an inverted Gaussian (Wald) distribution within the interval $(-1, 1)$. $\lambda$ is a hyperparameter controlling the strength of the regularization term, and $g(T) = abs(T/T_{max})$ normalizes and takes the absolute value of $T$. Notably, the second term represents the log-likelihood of the absolute value of $T$ under the inverted Gaussian distribution. 

\begin{figure}[t!]
    \centering  
    \includegraphics[width=0.99\linewidth]{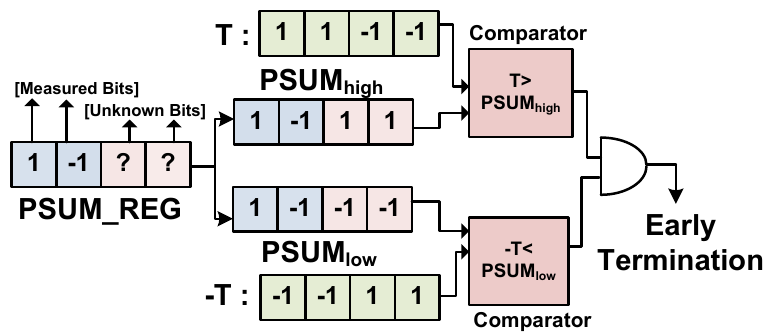} 
     \caption{\textbf{Digital Implementation of Early Termination:} This diagram shows the implementation of early termination using digital components.}
    \label{fig:Early_Termination Algo}
\end{figure}

Fig. \ref{fig: Early Termination Process}(a) illustrates the impact of the early termination technique on the distribution of the soft-thresholding parameter ($T$). In the absence of early termination, the distribution of $T$ is uniform. However, when we apply the unique loss function defined in Eq. (6), the $T$ parameter is driven towards -1 and 1, which aids output sparsification and workload reduction. Fig. \ref{fig: Early Termination Process}(b) demonstrates an example scenario showcasing how the comparison bounds on product sum (PSUM), i.e., $\text{PSUM}_{low}$ and $\text{PSUM}_{high}$ determining the early termination progressively tighten to improve the early termination opportunities with increasing bitplane processing. Fig. \ref{fig:Early_Termination Algo} shows a digital implementation of predictive early termination. As PSUM computations progress from MSB to lower significance bits, the unknown values are clamped to the highest and lowest values (1 or -1) to determine $\text{PSUM}_{low}$ and $\text{PSUM}_{high}$ for checking early termination criteria.     

Fig. \ref{fig: Early Termination Process}(c) shows the quantitative advantages of early termination. The figure shows the execution cycles for an 8-bit input vector processing case on 10,000 random cases. Without the early termination technique, the execution of all eight input bit planes will be necessary. However, the average number of input bit-planes is significantly lower with early termination while optimizing the distribution of $T$ parameters, as shown in Fig. \ref{fig: Early Termination Process}(a). For most operations, only one bit of extraction is sufficient before the operation can be predictably terminated. The average number of extraction cycles needed is less than 2 while maintaining the same level of accuracy. Fig. 10 shows the digital implementation of the early termination scheme, which can be implemented with digital comparators, shift registers, and logic.  

\section{Simulation Results and Discussions}
In this section, we discuss simulation results on the presented analog acceleration of frequency domain DNN processing. Simulations were conducted using HSPICE and predictive technology models for 16 nm CMOS technology \cite{6724591}.

\begin{figure*}[t!]
    \centering
    \includegraphics[width=0.24\linewidth]{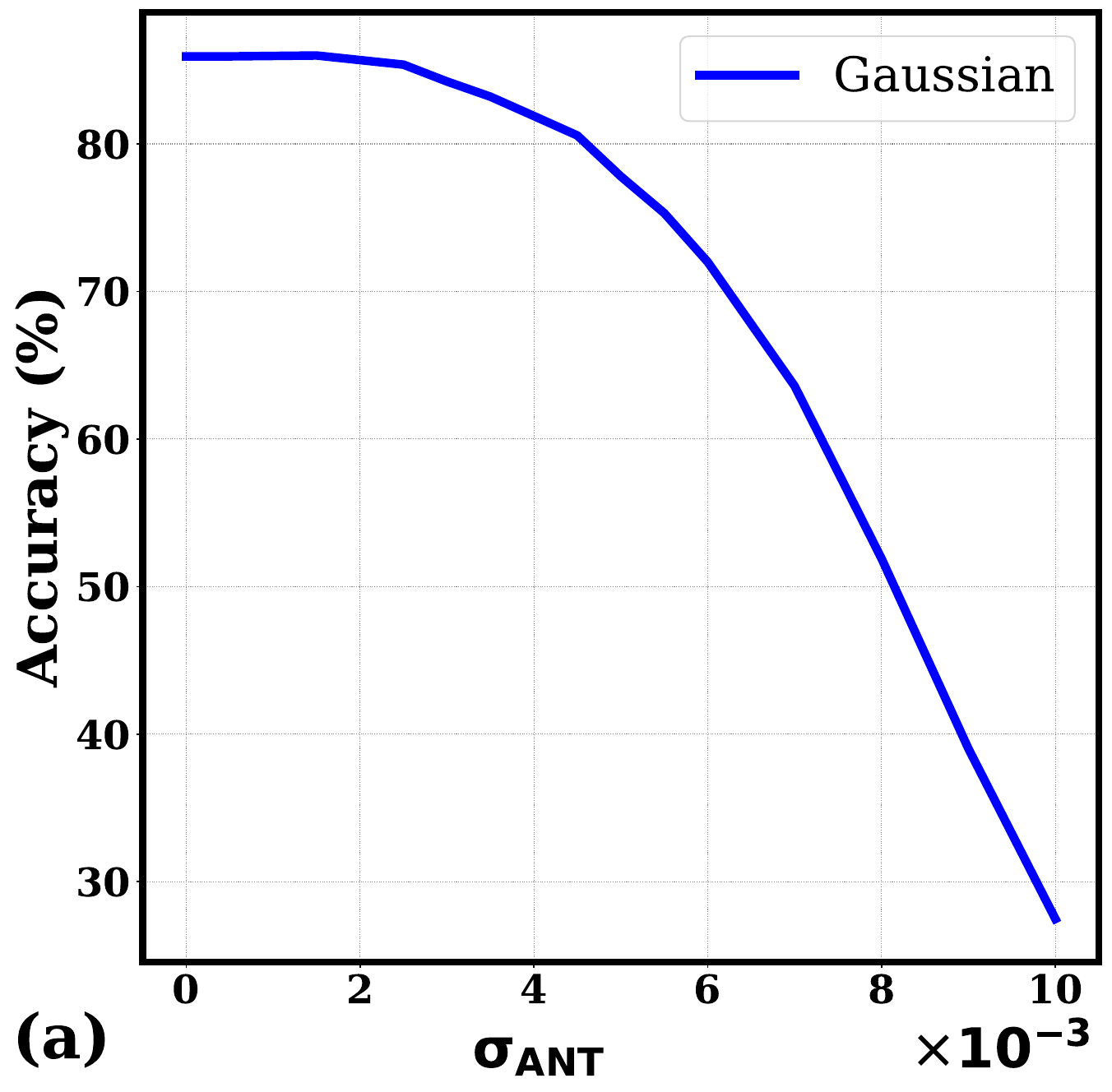}
    \includegraphics[width=0.24\linewidth]{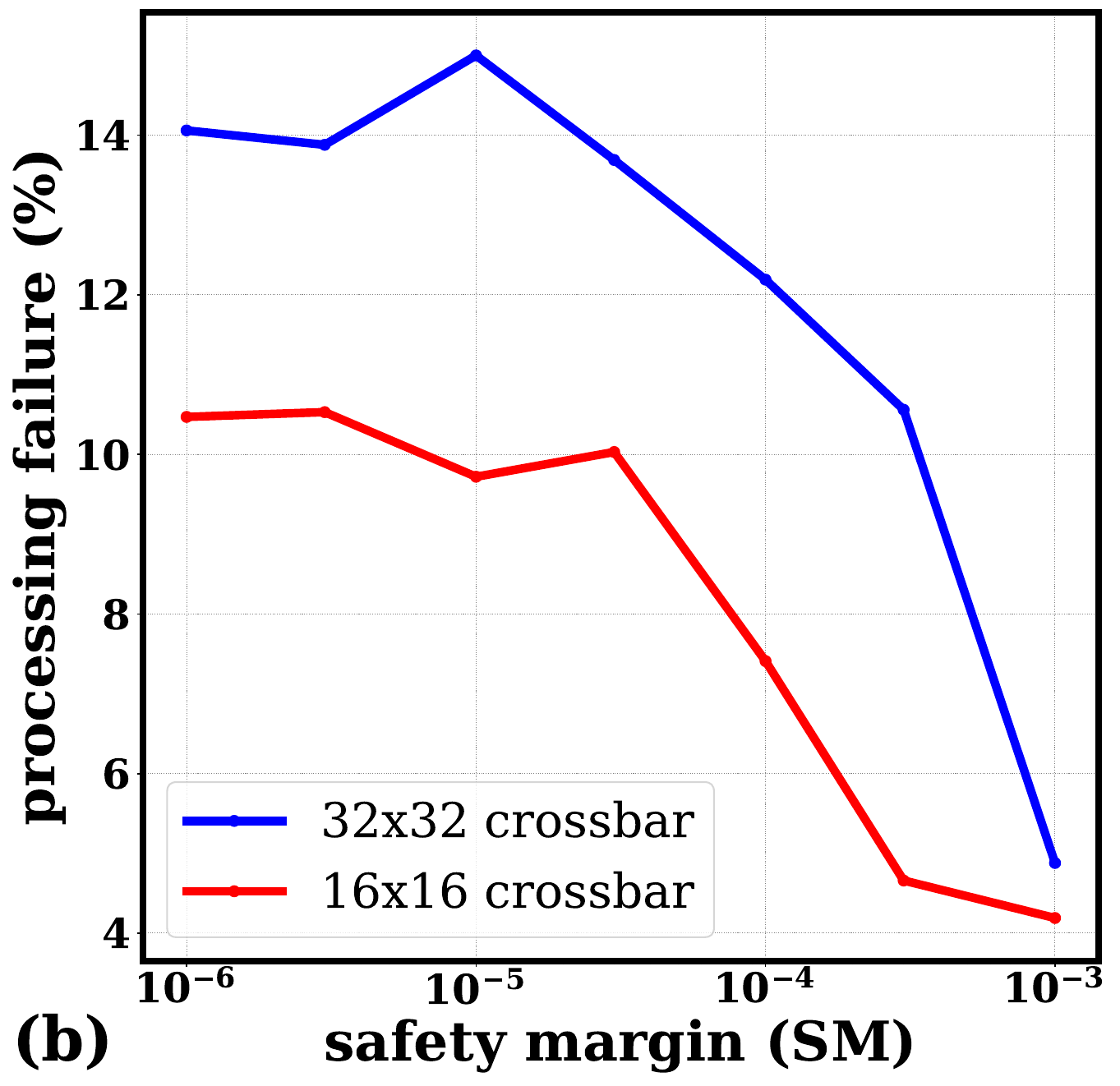}
    \includegraphics[width=0.2355\linewidth]{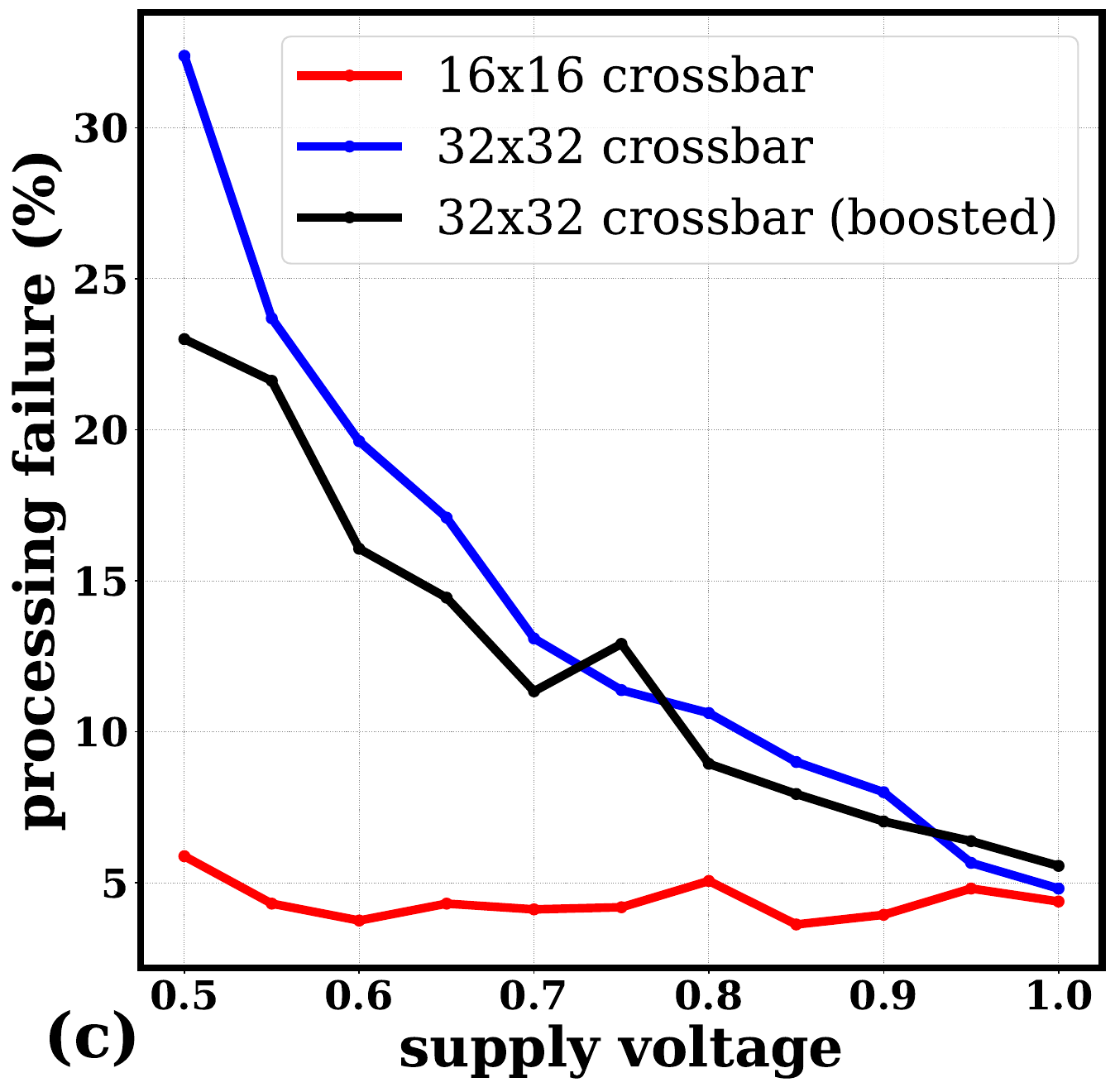}
    \includegraphics[width=0.241\linewidth]{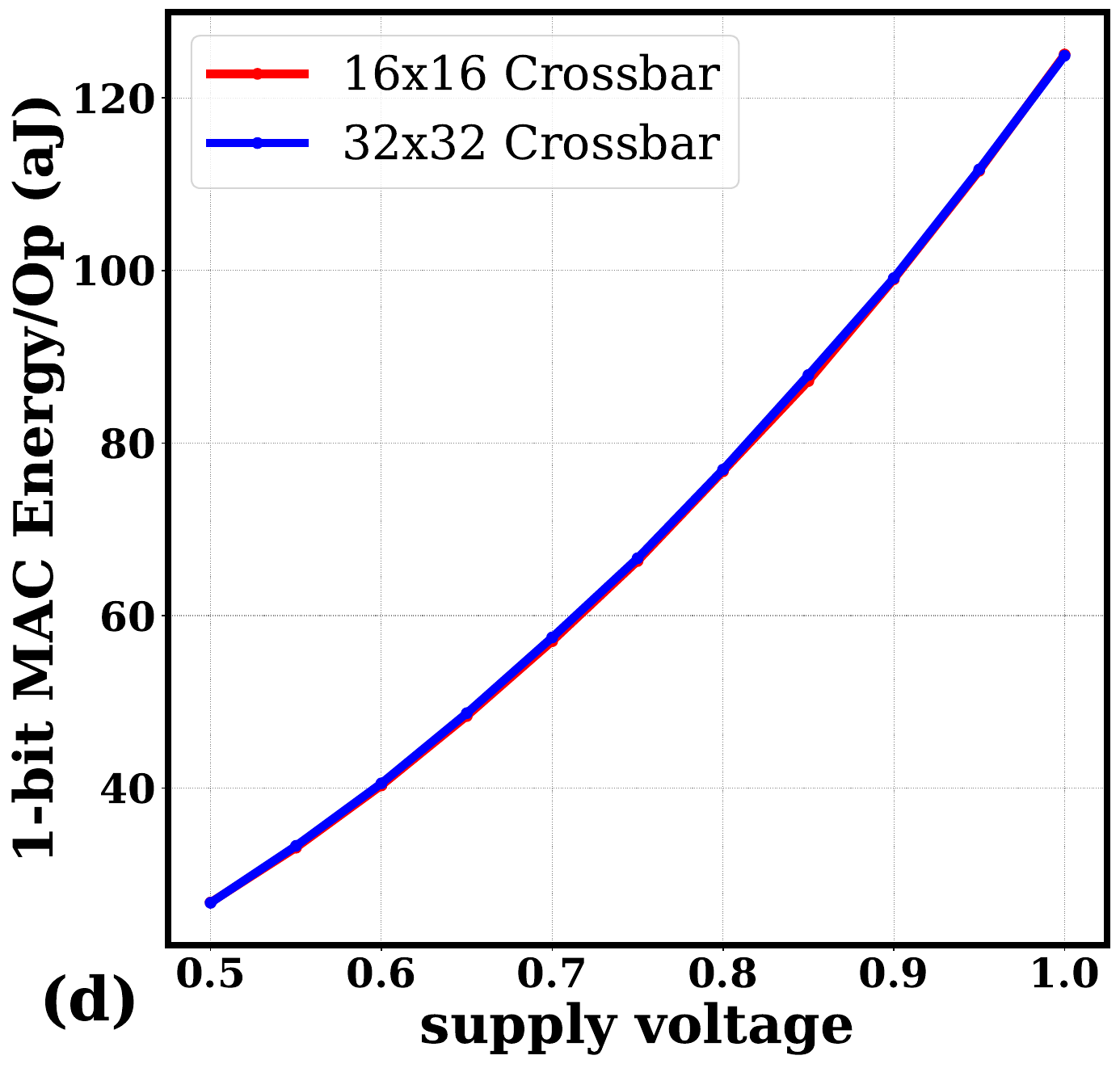}
    \caption{\textbf{Noise-Induced Quantization Effects:} \textbf{(a)} Algorithmic accuracy vs. standard deviation of noise added to normalized product sum ($\sigma_\text{ANT}$) (Algorithmic Noise Tolerance). \textbf{(b)} Processing failure vs. safety margin (SM) for $16 \times 16$ and $32 \times 32$ crossbars. The evaluations are conducted with a nominal supply voltage of 0.90 V.\textbf{(c)} Processing failure vs. supply voltage. \textbf{(d)} 1-bit MAC Energy/Operation (aJ) vs. supply voltage. }
    \label{fig:tolerance}
\end{figure*}

\begin{figure}[t!]
    \centering  
    \includegraphics[width=0.99\linewidth]{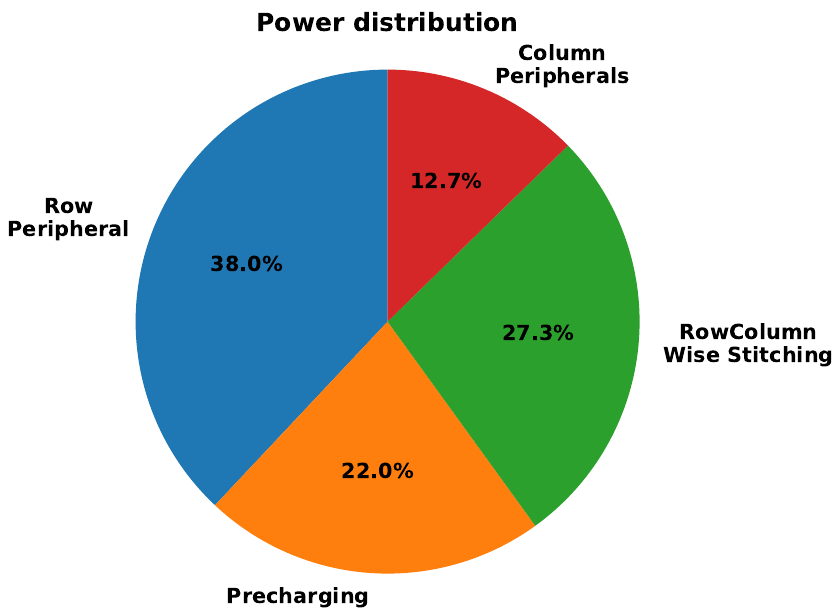} 
     \caption{\textbf{Power distribution:} This plot shows power distribution among various components of the operation for $16 \times 16$ crossbar.}
    \label{fig:Pie_chart}
\end{figure}

\subsection{Exploiting BWHT's Algorithmic Non-ideality Tolerance (ANT) for Processing Simplicity}
Due to the analog processing steps, computations in the proposed approach are susceptible to process variability and noise. For example, in Fig. 6, processing accuracy is susceptible to the offset and noise when the charge-domain processed PSUM is thresholded to -1 or 1 using analog comparators. Although sophisticated analog techniques such as autozeroing \cite{6724591}, chopper stabilization \cite{1051666}, layout guard rings \cite{7458868}, \textit{etc.} can minimize the non-idealities, they invariably come with the additional power or area overheads. In Fig. \ref{fig:tolerance}(a), we investigate the algorithmic non-ideality tolerance (ANT) of BWHT-based frequency domain processing, which can be leveraged to retain processing simplicity. The figure shows the accuracy trends of BWHT-based RESNET20's processing on CIFAR10 while injecting noise in the product sum. Here, we algorithmically mimic the bitplane-wise processing as implemented by our ADC/DAC-free analog structure in Fig. 4. Before the product sum (PSUM) output is digitized, noise is injected into it to emulate various non-idealities in the design as $\text{PSUM} \leftarrow \text{PSUM} + N(0, L_I\times \sigma_\text{ANT})$. $N()$ is a Gaussian random number generation function. $\sigma_\text{ANT}$ controls the standard deviation of injected noise. $L_I$ is the input vector length for which PSUM is computed. Fig. \ref{fig:tolerance}(a) shows the prediction accuracy with increasing $\sigma_\text{ANT}$ showing that $\sigma_\text{ANT} < 2\times10^{-3}$ causes an inconsequential impact on the overall accuracy.

In Fig. \ref{fig:tolerance}(b), we investigate the implications of such ANT of frequency domain processing to the simplicity of proposed analog acceleration. For a hundred random input and weight vectors, we compare the accuracy of the quantized product sum vector, as processed by our ADC/DAC-free analog structure in Fig. 4 under process variability, to the true values. The mismatch in the threshold voltage (i.e., local variability) is simulated by accounting $\sigma_{TH}$ = 24 mV for the minimum-sized transistors \cite{6724591} and scaling $\sigma_{TH}$ using Pelgrom's law for larger transistors. All analog cell transistors are minimum-sized. Peripherals are scaled according to the array size for the necessary driving strength. In the figure, if the true value of PSUM lies within the safety margin (SM), i.e., $|\text{PSUM}| < L_I \times \text{SM}$, its quantization errors are ignored considering BWHT's ANT in Fig. \ref{fig:tolerance}(a). $L_I$ is the input vector length mapped onto the analog processing array. Otherwise, product sum vector bit inaccuracy is marked as a processing failure. Fig. \ref{fig:tolerance}(b) shows the processing failure for 16$\times$16 and 32$\times$32 array sizes at sweeping safety margin (SM) in the above simulation setting. Notably, while in Fig. \ref{fig:tolerance}(a), $\sigma \sim 2\times10^{-3}$ causes an inconsequential impact on the overall accuracy of BWHT-based processing, with a comparable SM, the analog processing is accurate on more than 95\% on the considered random cases and array sizes. Thereby, leveraging BWHT-based processing algorithmic non-ideality tolerance allows processing simplicity of our analog acceleration scheme. 

\subsection{Design and Operating Space Exploration}
In Fig. \ref{fig:tolerance}(c), we demonstrate the variation in processing failure across different supply voltages for $16 \times 16$ and $32 \times 32$ crossbars. Evidently, for $32 \times 32$ crossbar, the processing failure increases sharply with lower supply voltage, whereas $16 \times 16$ crossbar design shows significantly better scalability with barely any increase in processing failure. Because of the row and column-wise parallelism enabled by stitching cells through column-merge and row-merge signals, a larger array becomes quadratically more vulnerable to process variability under supply voltage scaling. Importantly, in Fig. \ref{fig:tolerance}(c), boosting column-merge and row-merge signal voltages by 0.2 V reduces the processing failure for $32 \times 32$ crossbars. Still, the smaller crossbar allows an easier implementation by containing the processing variability even with a single VDD.

\begin{table*}
\small
    \setlength{\tabcolsep}{6pt}
    \renewcommand{\arraystretch}{1.2}
    \centering 
    \captionsetup{font=small}
    \caption{\textbf{Comparison between our approach and state-of-the-art for macro-level multiply-accumulate (MAC) processing}}
    \begin{tabularx}{\linewidth}{ccccccccc}
        \hline
        \textbf{Metric} & \textbf{Ours} & \textbf{\cite{9830276}} & \textbf{\cite{9250531}} & \textbf{\cite{9063078}} & \textbf{ \cite{9731716}} & \textbf{\cite{9062985}} & \textbf{\cite{9580448}}\\
        \hline
        Technology & \textbf{16nm} & 28nm & 7nm & 22nm & 22nm & 7nm & 16nm \\
        Computing mode & \textbf{CMOS Analog} & Neuromorphic & CMOS CiM & ReRAM CiM & CMOS Analog & CMOS CiM & CMOS Analog\\
        Weight Bits & 1 & 4 & 4 & 4 & ternary & 4 & 4\\
        Input Bits & \textbf{8} & 4 & 4 & 2 & 7 & 4 & 4\\
        Output Bits & \textbf{8} & 8 & 4 & 10 & -- & 4 & --\\        
        DAC & \textbf{No} & Capacitor & No & No & 7-bit & No & No\\
        ADC & \textbf{No} & No & 4-bit & No & 6-bit & 4-bit & 8-bit\\
        Neural Network & \textbf{MobileNetV2} & ResNet-18 & VGG9 & ResNet20 & ResNet20 & MLP & VGG\\
        Dataset & \textbf{CIFAR-10} & CIFAR-10 & CIFAR-10 & CIFAR-10 & CIFAR-10 & MNIST & CIFAR-10\\
        Accuracy & \textbf{91.04\%} & 92.80\% & 88.9\% & 90.18\% & 89\% & 98.47\% & 91.51\%\\
        TOPS/W & \textbf{1602}$^*$,\textbf{5311}$^{**}$ & 124.15 & 351 & 45.52 & 600 & 351 & 121\\        
        \hline
    \end{tabularx}
    \begin{flushleft}
        $^*$Energy efficiency is estimated without early termination strategy at VDD = 0.8 V, requiring eight cycles to process eight-bit input.\\
        $^{**}$Energy efficiency is estimated without early termination strategy at VDD = 0.8 V, which reduces the average number of bit-wise processing cycles to 1.34 for eight-bit input. The overheads of digital circuits for early termination are estimated from \cite{xie2015performance}.\\
        
    \end{flushleft}

\end{table*}

In Fig. \ref{fig:tolerance}(d), energy-per-operation for multiply-accumulate (MAC) processing for one bit-plane of the input vector's processing against frequency domain parameters are compared at varying supply voltage (VDD) for $16 \times 16$ and $32 \times 32$ crossbars. Notably, by splitting the bit lines of the analog crossbar cell-wise, the processing energy per operation is expected to be only weakly dependent on the crossbar size. This is also evident in the figure. Another notable aspect of our ADC/DAC-free crossbar design is that the simplicity of peripherals is leveraged for crossbar size downscaling. In typical analog crossbar processing, excessive crossbar downscaling results in more pronounced (area/power) resources dedicated to peripherals, thereby limiting downscaling. Meanwhile, smaller crossbars have lower bit line parasitic, and with suitable architecture choices, they can also better adapt to mapping deep learning layers of diverse sizes and channel widths. While the traditional analog crossbar approaches cannot take full advantage of crossbar downscaling, the proposed approach leads to better energy efficiency and allows better mapping by simplifying the peripherals. In Fig. \ref{fig:Pie_chart}, the average power consumption across various components for a $16 \times 16$ crossbar is demonstrated. Row/Column-wise stitching in our scheme incurs $\sim$27\% power overhead. However, matrix-level parallelism enabled by the scheme also improves the throughput of our design. Note that the scheme processes all elements of an input vector in parallel and computes all elements of the resulting output vector in parallel.    

\subsection{Comparison to State-of-the-Art}
Table I presents a comparison of our method with the current state-of-the-art techniques for macro-level multiply-accumulate (MAC) processing. Using 16$\times$16 crossbars and 8-bit input processing, our proposed method achieves an energy efficiency of 1602 TOPS/W without employing an early termination strategy. This efficiency increases to 5311 TOPS/W with the inclusion of the early termination strategy at VDD = 0.8 V. In the absence of early termination, processing an eight-bit input requires eight 1-bit MAC cycles, as illustrated in Fig. 6. However, by leveraging our early termination strategy, the average cycle count is significantly reduced to approximately 1.34, as highlighted in Fig. 9(c). The redesigned loss function achieves this efficiency by driving the $T$ parameters closer to 1 or -1, increasing the opportunities for early termination. It's worth noting that while scaling weight-bit precision often compromises the accuracy and energy efficiency in conventional deep learning implementations, our method effectively navigates around this challenge. Our approach processes inputs using only a one-bit frequency domain transformation matrix while applying high-precision parameters in the activation function, thus effectively sidestepping the typical trade-offs encountered during the model design phase. This synergistic integration of processing and model design facilitates simpler processing without compromising accuracy. 

\section{Conclusions}
We have introduced an ADC/DAC-free analog acceleration approach for deep learning by processing weight-input matrix products in the frequency domain. The proposed approach significantly reduces the necessary network size while matching the accuracy of a full-scale network. Despite the downsizing of weight matrices, the proposed approach maintains the structure of processing matrices by pruning in the frequency domain, thereby maximally leveraging the vector processing abilities for high performance, unlike in unstructured pruning-based approaches. This is due to the implementation of convolutions as element-wise multiplications in the transform domain. Analog computations in the proposed approach leverage physics for computing to minimize the workload while avoiding ADC/DAC for design simplicity and scalability of the processing node. Furthermore, we have implemented a predictive early termination strategy that intelligently terminates computations by accounting for the output sparsity due to rectified activation. To further enhance the potential of the early termination strategy, we have developed a novel loss function that optimizes model parameters to minimize the workload. With 16$\times$16 crossbars and 8-bit input processing, our proposed method delivers an energy efficiency of 1602 TOPS/W, which rises to 5311 TOPS/W when incorporating an early termination strategy at VDD = 0.8 V.

\bibliographystyle{IEEEtran}

\bibliography{main.bib}
\end{document}